\documentclass[11pt,a4paper]{article}
\pdfoutput=1
\usepackage{jheppub}
\usepackage{graphicx,psfrag}
\usepackage{bm}
\usepackage{mathbbol,verbatim}
\usepackage{slashed}
\usepackage{graphics}
\usepackage{color,ulem, amsmath, amssymb}
\usepackage{tikz}
\usetikzlibrary{trees}
\usetikzlibrary{decorations.pathmorphing}
\usetikzlibrary{decorations.markings}
\usetikzlibrary{decorations, decorations.markings, decorations.pathmorphing, arrows, graphs, shapes.geometric, snakes}
\usetikzlibrary{arrows}
\usepackage[colorlinks=true,linkcolor=blue,citecolor=blue]{hyperref}%
\allowdisplaybreaks

\definecolor{greeen}{rgb}{0.03,0.84,0.13}
\definecolor{test}{rgb}{0.03,0.74,0.33}
\definecolor{viol}{rgb}{0.44,0,0.94}
\definecolor{or}{rgb}{0.95,0.65,0}

\begin{document}
\begin{flushright}{{UMD-PP-019-02}}
\end{flushright}

\title{Ameliorating Higgs Induced Flavor Constraints on TeV Scale $W_R$}


\author[a]{Rabindra N. Mohapatra,}
\author[a]{Guanwen Yan,}
\author[b,c]{Yongchao Zhang}
\affiliation[a]{Maryland Center for Fundamental Physics, Department of Physics, University of Maryland, College Park, MD 20742, USA}
\affiliation[b]{Department of Physics and McDonnell Center for the Space Sciences,  Washington University, St. Louis, MO 63130, USA}
\affiliation[c]{Center for High Energy Physics, Peking University, Beijing 100871, China}

\date{\today}

\abstract{In the TeV scale minimal left-right symmetric model (LRSM) for neutrino masses, there is a tension between the flavor changing Higgs effects which prefer an $SU(2)_R$ breaking scale $v_R \gtrsim (15-25)$ TeV depending on whether the theory is kept invariant under charge conjugation ($Q_L\to (Q_R)^c$) or under parity ($Q_L\to Q_R$) respectively and an LHC accessible few-TeV range mass of $W_R$ boson which would require $v_R \lesssim 10 \, (15)$ TeV if $g_R/g_L = 1 (0.65)$. This requires one quartic coupling in the scalar potential to go non-perturbative, posing a theoretical problem if the $W_R$ is discovered at LHC.  We propose a simple extension of the minimal LRSM that adds a $B-L=0$ scalar triplet and study how this can  ameliorate this tension. We find that such a model is also constrained from various considerations and implies a lower bound on the $W_R$ mass of 8.1 (5.26) TeV  for the parity case with $g_R/g_L= 1\, (0.65)$ and 4.85 (3.16) TeV for the case of charge conjugation, if the flavor constraints have to be avoided while keeping all couplings perturbative. These mass ranges are accessible at the high-luminosity LHC. The model  also implies new decay mode of $W_R$ to two scalars which is absent in the minimal LRSM. Finally we comment on the impact of such a scalar multiplet for a class of dark matter extension of LRSM discussed in the literature recently. }

\maketitle

\section{Introduction}
The TeV scale left-right symmetric models (LRSMs)~\cite{LR1, LR2, LR3, LR4} based on the gauge group $SU(2)_L\times SU(2)_R\times U(1)_{B-L}$ have been widely discussed as the minimal extension of the standard model (SM)  that  accommodates small neutrino masses~\cite{MS} via the seesaw mechanism~\cite{seesaw1, MS, seesaw3, seesaw4, seesaw5}. The reason is that the two basic ingredients of seesaw mechanisms, i.e.  right handed neutrinos (RHNs) and their Majorana masses arising from $B-L$ breaking, are automatic in the LRSMs and do not have to be put in as additional inputs. An important practical question is whether the heavy $W_R$ boson predicted by LRSMs is detectable at the Large hadron collider (LHC) or one needs to go to higher energy colliders. For this purpose, one needs to know whether an LHC accessible $W_R$ with mass generally in the ($5-6$) TeV range~\cite{Ferrari:2000sp, Nemevsek:2018bbt, Chauhan:2018uuy} is compatible with low energy observations, e.g. the flavor changing  processes such as $K-\bar{K}$, $B_{d,s}-\bar{B}_{d,s}$ and $D-\bar{D}$ mixings induced by the new features of the model. One particular aspect that we focus in this paper concerns the implications of the scalar sector of LRSMs. In its minimal version~\cite{MS} which is widely considered in the literature~\cite{Gunion:1986im, Gunion:1989in, Deshpande:1990ip, Polak:1991vf, Barenboim:1996pt, Barenboim:2001vu, Azuelos:2004mwa, Jung:2008pz, Bambhaniya:2013wza, Dutta:2014dba, Bambhaniya:2014cia,  Crivellin:2018ahj, Maiezza:2015lza, Bambhaniya:2015wna, Dev:2016dja, Chakrabortty:2016wkl, Nemevsek:2016enw, Maiezza:2016ybz, Dev:2016vle, Dev:2017dui, Dev:2018upe, Borah:2018yxd, Dev:2018kpa, Dev:2018foq}, there is a bidoublet field $\Phi({\bf 2},{\bf 2},0)$ which couples to the SM quarks and leptons and gives them masses (as well as generate the Dirac masses for seesaw mechanism). The bidoublet field consists of two SM doublets with the second one being the parity partner of the first. This can be thought of as a two Higgs doublet extension of the SM (2HDM) except that the extra $SU(2)_R$ symmetry of the model constrains the couplings of $\Phi$ to quarks and leptons in a specific way. This leads to fewer free Yukawa coupling parameters than in a generic 2HDM. In fact, in a general 2HDM there are four Yukawa coupling matrices involving the up and down sectors of the quarks, whereas in the case of LRSM there are only two matrices given in the equation below:
\begin{eqnarray}
\label{eqn:Yukawa0}
{\cal L}^\phi_Y~=~h_{ij}\bar{Q}_{Li}\Phi Q_{Rj}+\tilde{h}_{ij}\bar{Q}_{Li}\tilde{\Phi} Q_{Rj}~+~h.c.
\end{eqnarray}
where $\tilde{\Phi}=\sigma_2\Phi^\ast \sigma_2$ ($\sigma_2$ being the second Pauli matrix), and $Q_L$ and $Q_R$ are respectively the left and right-handed quark doublets.
This property leads to the generation of large new scalar induced flavor changing neutral current (which we call FCNH) effects unlike the 2HDMs where it could be tuned to be zero. To see this heuristically, we can ignore CP violation, and note that the matrices $h_{ij}$ and $\tilde{h}_{ij}$ in Eq.~(\ref{eqn:Yukawa0}) are hermitian matrices due to left-right (LR) symmetry. One of the two matrices can be diagonalized by choice of basis without loss of generality. In this basis there are 9 free parameters (ignoring CP phases) describing the quark masses and mixings and they are all fixed by the six quark masses and three CKM mixing angles. Looking at the $\Phi$ field, we see that there are two neutral scalar fields, with the real part of the first one being dominantly the SM Higgs $h$. The couplings of the second neutral scalar field (denoted by $H_1+iA_1$ below)  to quarks are now fixed by quark masses and CKM angles. In the mass basis, it involves change of flavor due to the CKM rotations. It is this property that leads to large flavor changing effects from tree-level exchange of the new neutral scalar fields $H_1$ and $A_1$ (called here the FCNH effects) and puts lower bounds on the mass of these neutral scalar fields to be consistent with observations~\cite{Ecker:1983uh, Zhang:2007da, Maiezza:2010ic, Blanke:2011ry, Bertolini:2014sua}. This mass limits depend on the assumptions but can safely be anywhere from $\gtrsim 15$ TeV~\cite{Maiezza:2010ic} to $\gtrsim 25$ TeV~\cite{Zhang:2007da}, depending on whether one uses parity ($P$) which interchanges $Q_L \leftrightarrow Q_R$ (called LRP models) or the generalized charge conjugation ($C$) which interchanges $Q_L \leftrightarrow (Q_R)^c$ (called LRC models)  respectively~\cite{Maiezza:2010ic}. We will assume these values to be conservative, although they depend on assumptions. Since in the minimal LRSM these masses are given by  a formula $\sqrt{\alpha_3} v_R$, where $\alpha_3$ is a quartic coupling in the scalar potential (cf. Eq.~(\ref{eqn:potential})), these limits would imply that $v_R \gtrsim (15-25)$ TeV (for the LRC or LRP cases) if the coupling $\alpha_3$ is of order one. The latter implies that the $W_R$ boson mass given by $g_R v_R$ ($g_R$ being the gauge coupling for the gauge group $SU(2)_R$) is far above what LHC can access. Thus, if $W_R$ is discovered at the LHC, this would present a consistency problem for the minimal LRSM and would require its extension so that this tension does not exist. This kind of tension exists in both the type-I seesaw~\cite{MS} as well as inverse seesaw realization of the neutrino masses~\cite{Mohapatra:1986aw,Mohapatra:1986bd}  in LRSMs~\cite{Dev:2009aw, LalAwasthi:2011aa, Brdar:2018sbk}. In the bulk of this paper we focus on models with type-I seesaw; however, as we comment in Section~\ref{sec:conclusion}, our method can be applied to the inverse seesaw LRSMs as well.

In two papers, attempts were made to address this issue, in one case using a higher dimensional operator~\cite{Guadagnoli:2010sd} and in another using an extension that adds extra fermions and scalars~\cite{Mohapatra:2013cia}. In this paper, we provide a new economical extension of the LRSM by adding just a real $B-L=0$ $SU(2)_R$ scalar triplet $\delta_R$ and show that it provides a simple way to ameliorate this problem for a large range of $W_R$ mass. The addition of this field does not affect the neutrino mass features. Also, for the sake of simplicity, in this paper we have worked in the version of the model where parity is broken at a high scale~\cite{Chang:1983fu} so that the low energy spectrum does not contain the left-handed triplet $\Delta_L({\bf 3},{\bf 1},+2)$. The presence or absence of this field does not make any difference to the problem we are trying to address.

The addition of the $B-L=0$ triplet $\delta_R$ leads to several interesting results: (i) it increases the mass of $H_1$ and $A_1$ while keeping the coupling $\alpha_3$ perturbative i.e. $\alpha_3 \lesssim 1$; (ii) the presence of trilinear scalar couplings in the presence of the new triplet $\delta_R$ imposes further constraints on the model so that the FCNH solution can be maintained only if the mass of $W_R$ is larger than 5.26 TeV if $g_R/g_L=0.65$ for the LRP case, and 3.16 TeV for the LRC case, both of which are accessible at the LHC as well as high-luminosity LHC (HL-LHC); (iii) this new multiplet opens up a new decay channel for the $W_R$ to two scalar modes~\cite{Dobrescu:2015qna, Collins:2015wua}, all-be-it with a small branching ratio (BR), in  contrast with the minimal LRSM where it is absent; (iv) the presence of this new triplet scalar has also implications for dark matter (DM) extensions of the model providing more flexibility to the parameter space of the model.


The paper is organized as follows: In Section~\ref{sec:tension} we sketch briefly the level of tension in the minimal LRSM for various values of the gauge coupling ratio $g_R/g_L$. The masses and mixings among the neutral and singly-charged scalar fields are obtained in Section~\ref{sec:model}, in this section we explain how the new scalar triplet $\delta_R$ affects the FCNH constraints from the $K$ and $B$ mesons, and show the lower bounds on $W_R$ mass due to vacuum stability constraint arising from the 1-loop box diagrams. The singly-charged scalar $H_2^\pm$ from $\delta_R$ is rather interesting; in Section~\ref{sec:singlychargedscalar} we show how it is produced and decays at future hadron and lepton colliders. The effects of $H_2^\pm$ on heavy RHN decay is briefly commented in Section~\ref{sec:RHN}, the new scalar decay mode of $W_R$ boson is addressed in Section~\ref{sec:WR}, and the DM implications of the new scalar is detailed in Section~\ref{sec:DM}, before we comment and conclude in Section~\ref{sec:conclusion}. Some of the calculation details are collected in the appendices.


\section{Degree of the FCNH tension in the LRSM}
\label{sec:tension}

Before proceeding to discuss the model details, let us give more precisely the level of tension between the  LHC accessibility of $W_R$ and the FCNH constraints. Clearly the former depends on the value of the right-handed gauge coupling $g_R$ since that determines not only the mass of $W_R$ (for fixed right-handed scale $v_R$) but also the production rate of $W_R$ at the LHC.
The current LHC limits~\cite{Aaboud:2018spl, Sirunyan:2018pom} of $4.7$ TeV is for the special case of $g_L=g_R$ (for an analysis of LHC bounds on $W_R$, see e.g.~\cite{Mitra:2016kov}). The limits are relaxed if either the $V_{\rm CKM}$ in the left- and right-handed sectors are different~\cite{Langacker:1989xa, Barenboim:1996nd} or the RHN masses are larger than the $W_R$ mass~\cite{Frank:2018ifw}. Throughout this paper, we assume the left- and right-handed quark mixing matrices $V_{\rm CKM}^{(L,\,R)}$ are the same.

\begin{figure}[!t]
  \centering
  \includegraphics[width=0.42\linewidth]{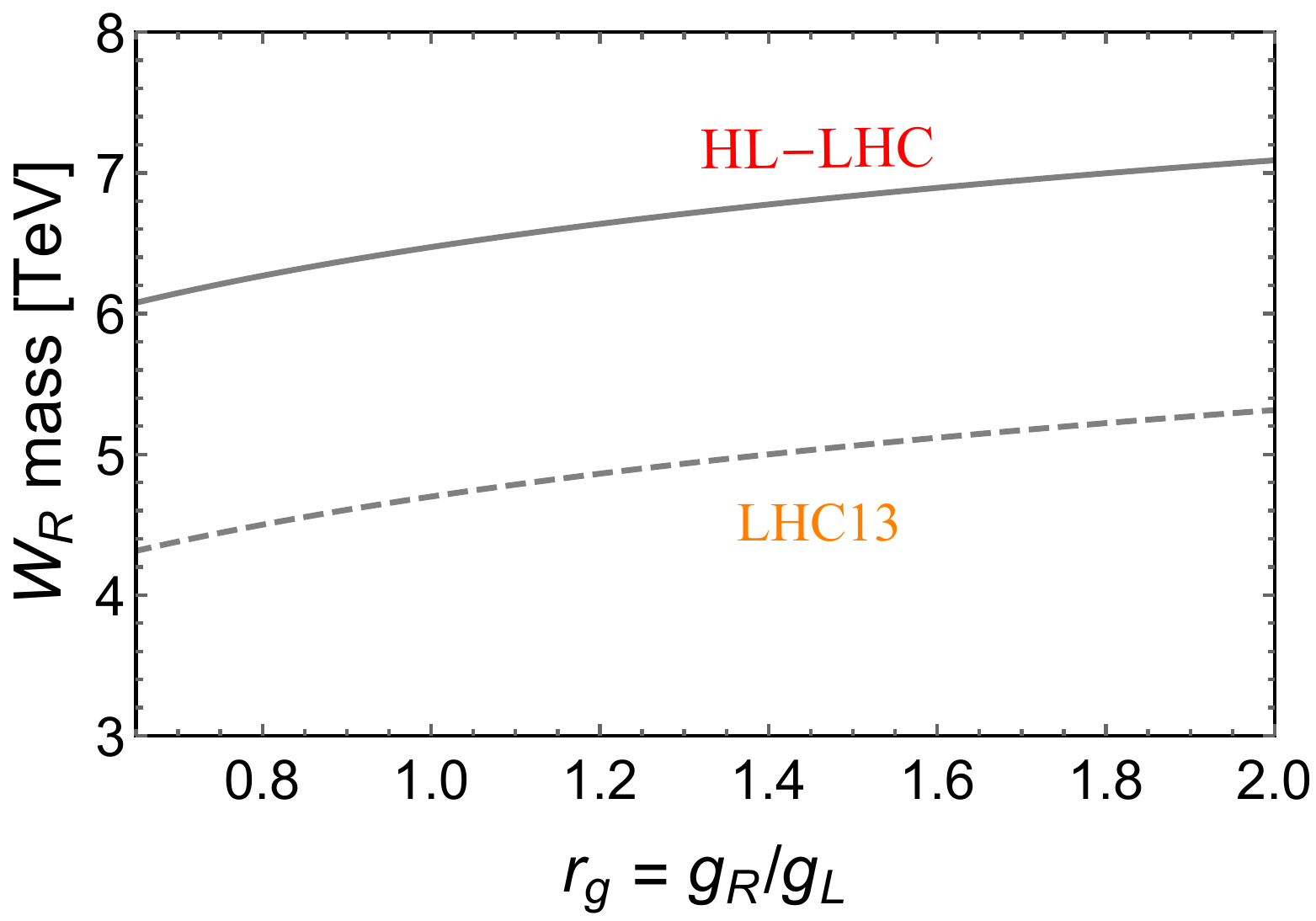}
  \caption{Variation of the LHC 13 TeV limits on $W_R$ mass (dashed)~\cite{Aaboud:2018spl, Sirunyan:2018pom} and the reach at the HL-LHC (solid) with $r_g=g_R/g_L$. This figure is from~\cite{Chauhan:2018uuy}. }
  \label{fig:WRreaches}
\end{figure}

As a result of the Majorana nature of the heavy RHNs $N$, the ``smoking-gun'' signatures of $W_R$ boson at hadron colliders are a pair of same-sign dilepton plus two jets without any significant missing energy, i.e. $W_R \to \ell^\pm N \to \ell^\pm \ell^\pm jj$ (here for simplicity we do not show explicitly all the flavor indices)~\cite{Keung:1983uu}. Given an integrated luminosity of 3000 fb$^{-1}$, the $W_R$ boson in the minimal LRSM can be probed up to about 6.5 TeV at the HL-LHC with center-of-mass energy of 14 TeV from the searches of same-sign dilepton signals if $g_R = g_L$. At leading order, the production cross section $\sigma (pp \to W_R) \propto g_R^2$, thus the $W_R$ prospect could go higher if $g_R > g_L$. In Fig.~\ref{fig:WRreaches} we show how this reach value changes as function of $g_R$. This figure is from Ref.~\cite{Chauhan:2018uuy} and shows both the current LHC 13 TeV limits from Refs.~\cite{Aaboud:2018spl, Sirunyan:2018pom} and the prospect at the HL-LHC. One should note that there is an absolute bound on the gauge coupling $r_g = g_R / g_L \gtrsim 0.55$~\cite{Brehmer:2015cia, Dev:2016dja}. If the gauge couplings are perturbative up to the grand unified theory (GUT) scale, the constraint on $g_R$ is more stringent, i.e. $g_R / g_L \gtrsim 0.65$~\cite{Chauhan:2018uuy}. As shown in Fig.~\ref{fig:WRreaches}, for a smaller $g_R$ with the value of $g_R / g_L = 0.65$, the $W_R$ discovery reach is about 6.1 TeV. This implies that the right-handed symmetry breaking scale $v_R \lesssim 10.0$ TeV if $g_L=g_R$, and somewhat larger for a smaller $g_R$, being $\lesssim 14.4$ TeV for $g_R/g_L = 0.65$.

On the other hand, the FCNH constraints imply $M_{H_1} \geq (15-25)$ TeV \cite{Zhang:2007da, Maiezza:2010ic},
which requires that in the minimal LRSM the quartic coupling $\alpha_3 \simeq M_{H_1}^2/v_R^2$ to be $\in [2.25,\, 6.25]$ for $v_R \simeq 10$ TeV, in the non-perturbative range. Furthermore, for an $\alpha_3 \gtrsim 1$, when the couplings in  the LRSM run up to higher energy scales, they would hit the Landau pole very quickly~\cite{Rothstein:1990qx, Chakrabortty:2013zja, Chakrabortty:2016wkl, Maiezza:2016ybz, Chauhan:2018uuy}. This creates a tension for the minimal LRSM for neutrino mass generation, not only  if the $W_R$ is discovered at LHC but also for a range of $W_R$ mass that is beyond the LHC accessible values. For the smallest value of $g_R\simeq 0.65 g_L \simeq 0.42$, $v_R\simeq 14.4$ TeV may be kinematically within the reach of HL-LHC, yet be compatible with FCNH constraints with a large quartic coupling $\alpha_3 \simeq 1.09$ if $M_{H_1} \simeq 15$ TeV. We also note that the limits on $W_R$ mass can be large from rare meson decays if the RHN mass is in the GeV range~\cite{Castillo-Felisola:2015bha, Mandal:2017tab}. We do not consider this case here.


\section{Extended scalar sector}
\label{sec:model}

In order to discuss how our extension of the LRSM helps with the FCNH tension, we analyze the scalar potential to get the mass spectra of neutral and charged scalar bosons in the new model. The scalar sector now consists of the following multiplets: the bidoublet field $\Phi({\bf 2},{\bf 2},0)$, the $SU(2)_R$ triplet field with $B-L=2$ denoted by $\Delta_R ({\bf 1},{\bf 3},+2)$ and the new real $B-L=0$ field $\delta_R ({\bf 1},{\bf 3},0)$ which is not present in the minimal LRSM. As noted earlier, this is the effective low energy version of the parity symmetric model where the parity symmetry has been broken at a high scale~\cite{Chang:1983fu} so that the $SU(2)_L$ triplet $\Delta_L({\bf 3},{\bf 1},+2)$ and $\delta_L$  are absent from the Lagrangian. The detailed field content of the multiplets are:
\begin{eqnarray}\label{fields}
\Phi = \left(\begin{matrix}\phi^0_1 & \phi^+_2\\\phi^-_1 & \phi^0_2\end{matrix}\right), \quad
  \Delta_R = \left(\begin{matrix} \frac{1}{\sqrt2} \Delta^+_R & \Delta^{++}_R \\ \Delta^0_R & - \frac{1}{\sqrt2} \Delta^+_R
\end{matrix}\right), \quad
\delta_R = \left(\begin{matrix}  \delta^0_R & \delta^{+}_R \\ \delta^-_R & -  \delta^0_R \end{matrix}\right)
 \end{eqnarray}
The most general scalar potential of the bidoublet field $\Phi$, the triplet field $\Delta_R$ and the real field $\delta_R$ is given by:
\begin{eqnarray}
\label{eqn:potential}
\mathcal{V} & \ = \ &
- \mu_1^2 \: {\rm Tr} (\Phi^{\dag} \Phi)
- \mu_2^2 \left[ {\rm Tr} (\tilde{\Phi} \Phi^{\dag}) + {\rm Tr} (\tilde{\Phi}^{\dag} \Phi) \right]
- \mu_3^2 \:  {\rm Tr} (\Delta_R \Delta_R^{\dag})
- \mu_4^2 \:  {\rm Tr} (\delta_R \delta_R^{\dag}) \nonumber \\
&& + M_2 {\rm Tr}(\Phi \delta_R \Phi^\dagger)
+ M_3{\rm Tr}(\Delta_R \delta_R\Delta_R^{\dag})\nonumber \\
&& + \lambda_1 \left[ {\rm Tr} (\Phi^{\dag} \Phi) \right]^2 + \lambda_2 \left\{ \left[
{\rm Tr} (\tilde{\Phi} \Phi^{\dag}) \right]^2 + \left[ {\rm Tr}
(\tilde{\Phi}^{\dag} \Phi) \right]^2 \right\} \nonumber \\
&&+ \lambda_3 \: {\rm Tr} (\tilde{\Phi} \Phi^{\dag}) {\rm Tr} (\tilde{\Phi}^{\dag} \Phi) +
\lambda_4 \: {\rm Tr} (\Phi^{\dag} \Phi) \left[ {\rm Tr} (\tilde{\Phi} \Phi^{\dag}) + {\rm Tr}
(\tilde{\Phi}^{\dag} \Phi) \right] \nonumber \\
&& + \rho_1  \left[ {\rm Tr} (\Delta_R \Delta_R^{\dag}) \right]^2
+ \rho_2 \: {\rm Tr} (\Delta_R \Delta_R) {\rm Tr} (\Delta_R^{\dag} \Delta_R^{\dag})
+ \eta_1  \left[ {\rm Tr} (\delta_R \delta_R^{\dag}) \right]^2 \nonumber \\
&&+ \alpha_1 \: {\rm Tr} (\Phi^{\dag} \Phi) {\rm Tr} (\Delta_R \Delta_R^{\dag})
+ \alpha_2 \left[  {\rm Tr} (\tilde{\Phi}^{\dag} \Phi) + {\rm Tr} (\Phi^{\dag} \tilde{\Phi})  \right]
{\rm Tr} (\Delta_R \Delta_R^{\dag})
+ \alpha_3 \: {\rm Tr}(\Phi^{\dag} \Phi \Delta_R \Delta_R^{\dag}) \nonumber \\
&&+ \beta_1 \: {\rm Tr} (\Phi^{\dag} \Phi) {\rm Tr} (\delta_R \delta_R^{\dag})
+ \beta_2 \left[  {\rm Tr} (\tilde{\Phi}^{\dag} \Phi) + {\rm Tr} (\Phi^{\dag} \tilde{\Phi})  \right] {\rm Tr} (\delta_R \delta_R^{\dag})
+ \gamma_1 \: {\rm Tr} (\Delta_R^{\dag} \Delta_R) {\rm Tr} (\delta_R \delta_R^{\dag}) \,. \nonumber \\ &&
\end{eqnarray}
For simplicity, we have assumed all the parameters in the potential are real. Minimizing the potential with respect to the VEVs $\langle \phi_1^0 \rangle = \kappa_1$, $\langle \phi_2^0 \rangle = \kappa_2$, $\langle \Delta^0_R\rangle=v_R$ and $\langle \delta^0_R\rangle =w_R$ leads us to the relations
\begin{eqnarray}
\frac{\mu_1^2}{v_R^2} & \ = \ &
- \frac{x M_2}{\sqrt2 v_R} + \alpha_1 + \beta_1 x^2 + {\cal O} (\varepsilon^2) \,, \\
\frac{\mu_2^2}{v_R^2} & \ = \ &
\alpha_2 + x^2 \beta_2 + {\cal O} (\varepsilon^2) \,, \\
\frac{\mu_3^2}{v_R^2} & \ = \ &
- \frac{x M_3}{\sqrt2 v_R} +
2\rho_1 + \gamma_1 x^2 + {\cal O} (\varepsilon^2) \,, \\
\frac{\mu_4^2}{v_R^2} & \ = \ &
- \frac{M_3}{2\sqrt2 x v_R} +
\gamma_1 + 2 \eta_1 x^2 + {\cal O} (\varepsilon^2) \,,
\end{eqnarray}
where we have defined $\varepsilon \equiv v_{\rm EW}/v_R$, $\xi \equiv \kappa_2 / \kappa_1$ and $x\equiv w_R / v_R$ with the electroweak VEV $v_{\rm EW} = \sqrt{\kappa_1^2 + \kappa_2^2} \simeq \kappa_1$ and the VEV ratios $\varepsilon,\, \xi \ll 1$. These relations can be used to determine the $\mu_i^2$ parameters in the potential Eq.~(\ref{eqn:potential}), with the trilinear coefficients $M_{2,\,3}$ left as free parameters.

\subsection{Neutral scalars and ameliorating FCNH}
\label{sec:neutralscalars}
Following~\cite{Dev:2016dja}, the mass matrix for the CP-even neutral scalars reads, up to the second order of the small parameters $\varepsilon$ and $\xi$ and in the basis of the real components $\{ \phi_1^{0\, {\rm Re}} ,\, \phi_2^{0\, {\rm Re}} ,\, \Delta_R^{0\, {\rm Re}} ,\, \delta_R^{0} \}$:
\begin{eqnarray}
\label{eqn:scalarmatrix}
{\cal M}^0 \ \simeq \
\left( \begin{matrix}
4 \lambda_1 \varepsilon^2  &
- \tilde{\alpha}_3 \xi & 2\alpha_1 \varepsilon & -[\frac{\tilde\alpha_3 - \alpha_3}{2x} - 2\beta_1 x ] \varepsilon \\
- \tilde{\alpha}_3 \xi & \tilde{\alpha}_3 & 4 \alpha_2 \varepsilon & 4 x \beta_2 \varepsilon \\
2\alpha_1 \varepsilon & 4 \alpha_2 \varepsilon & 4 \rho_1 &
- \frac{r}{\sqrt2} + 2\gamma_1 x \\
-[\frac{\tilde\alpha_3 - \alpha_3}{2x} - 2\beta_1 x] \varepsilon & 4 x \beta_2 \varepsilon &
- \frac{r}{\sqrt2} + 2\gamma_1 x &
\frac{r}{2\sqrt2 x} + 4 \eta_1 x^2
\end{matrix} \right) v_R^2 \,,
\end{eqnarray}
where we have defined the dimensionless parameters
\begin{eqnarray}
\label{eqn:tildealpha3}
\tilde\alpha_3 \ = \
\alpha_3 + \frac{\sqrt2 x M_2}{v_R} \ = \
\alpha_3 + \frac{\sqrt2 M_2 w_R}{v_R^2} \,, \quad
r \ = \ \frac{M_3}{v_R} \,.
\end{eqnarray}
The neutral scalar mass matrix in Eq.~(\ref{eqn:scalarmatrix}) can be diagonalized by the rotation matrix, up to the order of ${\cal O} (\varepsilon,\,\xi)$
\begin{eqnarray}
\label{eqn:scalarmixing}
\left( \begin{array}{c}
h \\ H_1 \\ H_3 \\ H_4 \\
\end{array} \right) =
\left(
\begin{array}{ccccc}
 1  & \xi  & -\sin\theta_1 & -\sin\theta_3   \\
 -\xi  & 1  & -\sin\theta_2 & -\sin\theta_4  \\
 \sin\theta_1 & \sin\theta_2 & \cos\theta_5  & -\sin\theta_5  \\
 \sin\theta_3 & \sin\theta_4  & \sin\theta_5 & \cos\theta_5
\end{array} \right)
\left( \begin{array}{c}
\phi_1^{\rm 0 \, Re} \\ \phi_2^{\rm 0 \, Re} \\ \Delta_R^{\rm 0 \, Re} \\ \delta_R^{\rm 0}  \\
\end{array} \right) \,,
\end{eqnarray}
where the mixing angles $\sin\theta_{1,2,3,4}$ are expected to be small and $\sin\theta_5$ is potentially large. The scalar $h$ is identified as the SM-like Higgs, with mass
\begin{equation}
\label{eqn:hmass}
m_h^2 \ \simeq \
\left[ 4\lambda_1 -\frac{\alpha_1^2}{\rho_1}
- \left( \frac{\tilde{\alpha}_3-\alpha^2}{2x} - 2\beta_1 x \right)^2 \left( \frac{r}{2\sqrt2 x} + 4 \eta_1 x^2 \right)^{-1} \right] \kappa_1^2 \,,
\end{equation}
where we have used
\begin{eqnarray}
\sin\theta_1 & \ \simeq \ &
\frac{\alpha_1 \varepsilon}{(2\rho_1)} \,, \nonumber \\
\sin\theta_3 & \ \simeq \ &
- \left( \frac{\tilde{\alpha}_3-\alpha^2}{2x} - 2\beta_1 x \right) \left( \frac{r}{2\sqrt2 x} + 4 \eta_1 x^2 \right)^{-1} \varepsilon
\end{eqnarray}
in getting this value. $H_1$ the neutral scalar predominantly from the real part of heavy doublet $\phi_2$, and $H_{3,4}$ mostly from the component $\Delta_R^{\rm 0\, Re}$ and $\delta_R^0$ of the right-handed triplets.

If the trilinear mass parameter $M_2 \gg v_R^2/w_R$, the mass of $M_{H_1}^2 \simeq \tilde\alpha_3 v_R^2$ is naturally much higher than the $v_R$ scale without a large quartic coupling $\alpha_3$ in the potential (\ref{eqn:potential}) as is required in the minimal LRSM. This is the key result that substantiates our claim that addition of $\delta_R$ field ameliorates the FCNH problem of LRSMs without compromising perturbativity of all scalar couplings. We will see in Section~\ref{sec:loop} that, when the loop corrections are taken into consideration, $\tilde\alpha_3$ can not be, however, too large without creating problems for the theory.

In the limit of $r \to 0$ and $\gamma_{1} \to 0$, the two heavy neutral scalars $H_{3,4}$ are almost purely from the components $\Delta_R^{\rm 0\, Re}$ and $\delta_R^0$, and the mixings of $h$ and $H_1$ with $H_3$ are the same as in the minimal model, which are respectively proportional to the quartic couplings $\alpha_{1,2}$ in the potential (\ref{eqn:potential}). If the couplings are comparable, i.e. ${\cal M}^0_{34} \ll {\cal M}^0_{33,\, 44}$ in the matrix (\ref{eqn:scalarmatrix}), the two states $\Delta_R^{\rm 0\, Re}$ and $\delta_R^0$ mix sizably with each other, and their mixing angle $\sin\theta_5$ in the matrix (\ref{eqn:scalarmixing}) would induce very rich phenomenology in the scalar sector (such as the long-lived $H_3$~\cite{Dev:2016vle,Dev:2017dui}) as well as for the searches of $W_R$ 
 in the LRSM (see Section~\ref{sec:WR}).

If the quartic couplings $\beta_{1}$ and the mass parameter $M_3$ are not very large, then the mixing of the SM-like Higgs with $\delta_R^{0\, {\rm Re}}$ is dominated by the $\tilde\alpha_3$ term in the element ${\cal M}^0_{14}$ of the mass matrix (\ref{eqn:scalarmatrix}), i.e. the trilinear $M_3$ term in the potential Eq.~(\ref{eqn:potential}), as Eq.~(\ref{eqn:tildealpha3}) implies that $M_3 \sim \tilde\alpha_3 v_R^2/w_R$.
More explicitly
\begin{eqnarray} \label{eqn:theta3}
\sin\theta_3 \ \simeq \
\frac{\tilde\alpha_3\varepsilon}{2x} /  4\eta_1 x^2 \ = \
\frac{1}{8\eta_1 }
\frac{v_{\rm EW}}{w_R}
\frac{M_{H_1}^2}{w_R^2} \,,
\end{eqnarray}
in other words, the mixing angle $\sin\theta_3$ is also related to the heavy doublet mass $M_{H_1}$ in addition to the VEV ratios $\varepsilon = v_{\rm EW}/v_R$ and $x = w_R/v_R$. There are very stringent FCNH constraints on the heavy doublet mass $M_{H_1} \gtrsim (15 - 25)$ TeV, then the current constraints of Higgs precision measurements on the mixing angle $\sin\theta_3$ set a lower bound on the $w_R$ scale,
\begin{eqnarray}
\label{eqn:wRlimit}
w_R \ \gtrsim \
\left( \frac{v_{\rm EW} M_{H_1}^2}{8\eta_1 \sin\theta_3} \right)^{1/3}
\ \simeq \ (2.8 \, {\rm TeV}) \times
\left( \frac{\sin\theta_3}{0.22} \right)^{-1/3}
\left( \frac{\eta_1}{1} \right)^{-1/3}
\left( \frac{M_{H_1}}{15 \, {\rm TeV}} \right)^{2/3} \,.
\end{eqnarray}
Here we have taken the current LHC constraints on a generic scalar mixing with the SM Higgs $\lesssim 0.22$~\cite{Falkowski:2015iwa}, which could be further improved up to 0.13 at future lepton colliders~\cite{Profumo:2014opa}. If the quartic coupling $\eta_1 < 1$ and the heavy scalar mass $M_{H_1} > 15$ TeV, the lower limit on $w_R$ will get more constraining as scaled in Eq.~(\ref{eqn:wRlimit}).

Now let us look at the mass spectrum of the imaginary part of $\phi^0_2$ which can also lead to large FCNH effects. For this purpose, we need to diagonalize the mass matrix involving this field. Since $\delta_R$ has zero $B-L$ charge, we can choose it to be a real field. The mass matrix of the imaginary part of the three neutral scalars is, in the basis of $\{ \phi_1^{0\, {\rm Im}} ,\, \phi_2^{0\, {\rm Im}} ,\, \Delta_R^{0\, {\rm Im}} \}$ and  in our approximation:
\begin{eqnarray}
\label{eqn:imscalarmatrix}
\left( \begin{matrix}
\tilde{\alpha}_3 \xi^2 & \tilde{\alpha}_3 \xi & 0 \\
\tilde{\alpha}_3 \xi & \tilde{\alpha}_3 & 0 \\
0 & 0 & 0
\end{matrix} \right) v_R^2 \,,
\end{eqnarray}
We can easily diagonalize the matrix:
\begin{eqnarray}
\label{eqn:imscalarmixing}
\left( \begin{array}{c}
G_Z \\ A_1 \\ G_{Z_R} \\
\end{array} \right) =
\left(
\begin{array}{ccccc}
 1  & -\xi  & 0  \\
  \xi & 1 & 0 \\
 0  & 0  & 1
\end{array} \right)
\left( \begin{array}{c}
\phi_1^{\rm 0 \, Im} \\ \phi_2^{\rm 0 \, Im} \\ \Delta_R^{\rm 0 \, Im} \\
\end{array} \right) \,,
\end{eqnarray}
where the two massless states $G_Z$ and $G_{Z_R}$ are eaten respectively by the $Z$ and $Z_R$ boson, leaving only one massive state $A_1$ with mass $M_{A_1}^2 = \tilde{\alpha}_3 v_R^2$. Again we see that the mass of the field $A_1$, which is predominantly from ${\rm Im}\,\phi^0_2$, is also given by the enhanced coupling $\tilde{\alpha}_3$ and can be made large without choosing any quartic coupling to be large. As in the minimal LRSM, the masses of $H_1$, $A_1$, as well as the singly-charged scalar $H_1^\pm$ which is predominately from $\phi_2^\pm$, are quasi-degenerate at the leading order of the small VEV ratios $\xi$ and $\varepsilon$.

\subsection{Loop corrections and restrictions on parameter space}
\label{sec:loop}

\begin{figure}[!t]
  \centering
  \includegraphics[width=0.32\textwidth]{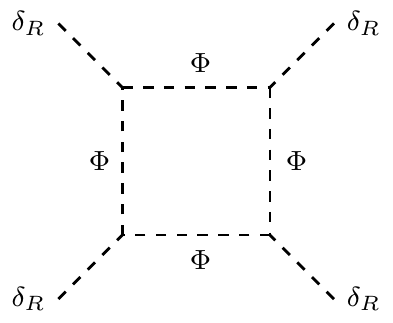}
  \caption{Feynman diagram for one-loop induced 4-$\delta_R$ interaction from the $M_2$ term in the potential~(\ref{eqn:potential}).}
\label{fig:4delta}
\end{figure}

\begin{figure}[!t]
  \centering
  \includegraphics[width=0.49\textwidth]{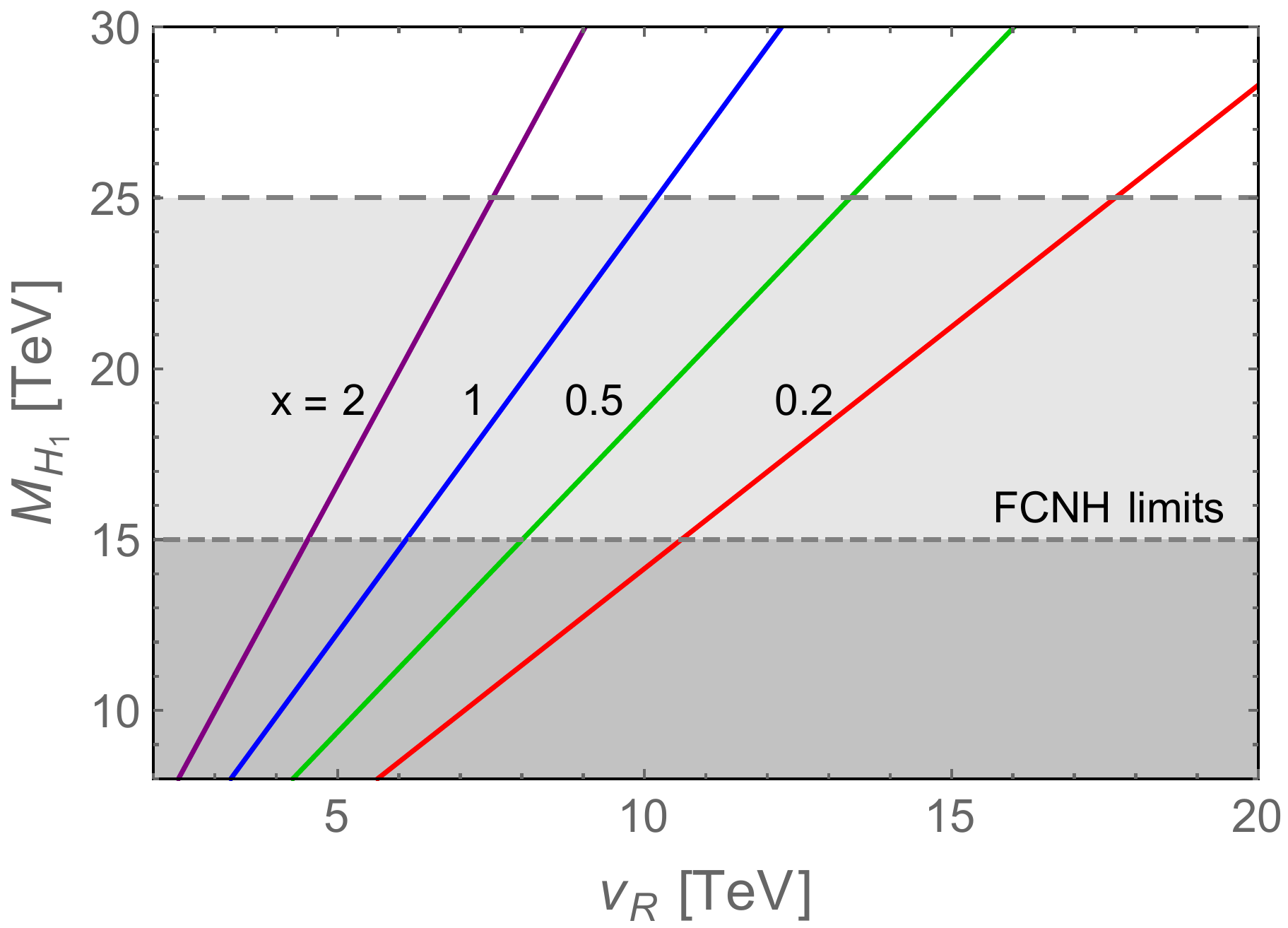}
  \includegraphics[width=0.49\textwidth]{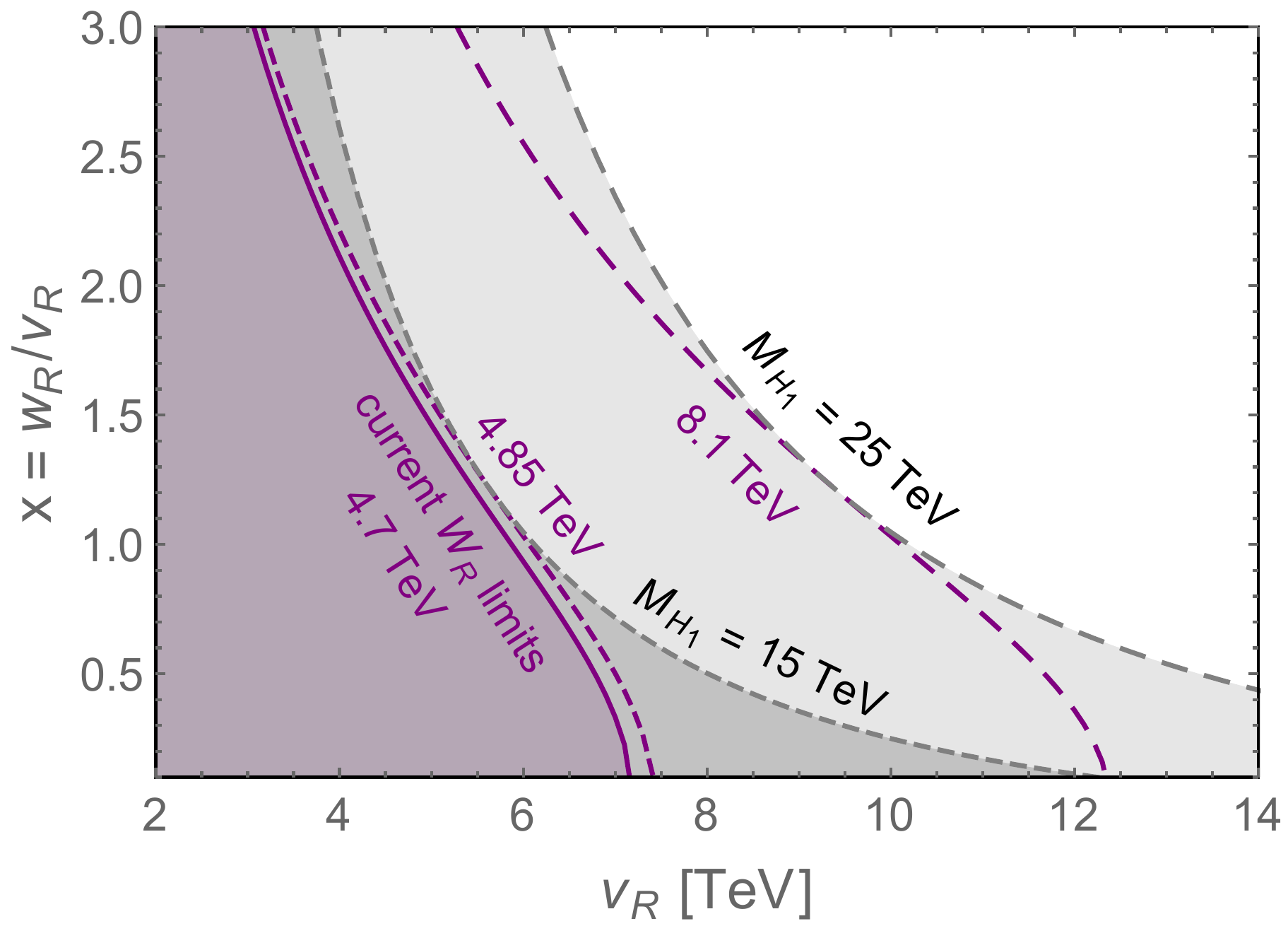}
  \caption{{\it Left panel}: the heavy bidoublet scalar mass $M_{H_1}$ as function of the VEV $v_R$ with respectively the ratio $x = w_R/v_R = 0.2$ (red), 0.5 (green), 1 (blue) and 2 (purple). The shaded regions are excluded by the FCNH limits of $M_{H_1} > (15 - 25)$ TeV (short and long dashed gray lines). {\it Right panel}: the same data sets of the FCNH limits as in the left panel, where we also show the current $W_R$ mass limit of 4.7 TeV (solid purple) assuming the gauge coupling $g_R = g_L$, as well as the contours of $M_{W_R} = 4.85$ TeV (short-dashed purple) and 8.1 TeV (long-dashed purple). See text for more details. We have taken $\alpha_3 = \eta_1 = 1$ in Eq.~(\ref{eqn:MH1}) in the plots.}
  \label{fig:MH1}
\end{figure}

Because of the $M_2$ trilinear term in the scalar potential (\ref{eqn:potential}), there is a loop induced 4-$\delta_R$ interaction in our model which does not exist in the minimal LRSM. The effective Lagrangian of this interaction (see Fig.\ref{fig:4delta}) at the one-loop level is found to be
\begin{equation}
{\cal L}_{\rm loop} \ = \
\frac{M_2^4}{16\pi^2v_R^4}\left[{\rm Tr}(\delta_R\delta_R^{\dagger})^2\right]
\end{equation}
In the potential (\ref{eqn:potential}), there is already a quartic term of $\delta_R$, i.e. the $\eta_1$ term.
The sum of these two terms must be less than zero,
otherwise potential will go to $-\infty$ for large $\delta_R$ and the vacuum will be unstable~(For use of similar argument in other models, see e.g.~\cite{Babu:2002uu, Babu:2013yca}). This imposes an upper bound on the mass parameter
\begin{equation}
M_2 \ < \ \sqrt{4\pi} \eta_1^{1/4} v_R \,.
\end{equation}
Plugging this into the approximate expression of $H_1$ mass, we get
\begin{eqnarray}
\label{eqn:MH1}
M^2_{H_1^0} & \ \approx \ & \tilde{\alpha}_3v_R^2
\ = \ \alpha_3v_R^2+\sqrt{2}M_3w_R \nonumber \\
& \ < \ & (\alpha_3 + 2\sqrt{2\pi} \eta_1^{1/4} x) v_R^2
\ \simeq \ (\alpha_3 + 5 \eta_1^{1/4} x) v_R^2 \,.
\end{eqnarray}
The FCNH constraints on $H_1$ mass therefore impose severe constraints  on the VEVs $v_R$ and $w_R$ (or equivalently on the ratio $x = w_R/v_R$). Setting the quartic couplings $\alpha_3 = \eta_1 = 1$, some contours of $M_{H_1}$ have been shown in the left panel of Fig.~\ref{fig:MH1}, as functions of $v_R$ and $x = 0.2$ (red), 0.5 (green), 1 (blue) and 2 (purple). The darker (lighter) shaded regions are excluded by the FCNH limit of 15 (25) TeV on the $H_1$ mass. This implies a lower bound on the $W_R$ mass, which is shown in the right panel of Fig.~\ref{fig:MH1}. In the extended LRSM we are considering, the $W_R$ boson mass reads
\begin{equation}
M_{W_R}^2 \ = \ g_R^2 v_R^2 \left( 1 + \frac12 x^2 \right) \,,
\end{equation}
which is also a function of $v_R$ and $x$, up to the gauge coupling $g_R$. If $g_R = g_L$, then the FCNH limit of 15 (25) TeV implies that $M_{W_R} > 4.85 \, (8.1)$ TeV, depicted as the short and long dashed purple curves in the right panel of Fig.~\ref{fig:MH1}. This is significantly higher than the current LHC limit of 4.7 TeV on the $W_R$ mass for $g_R = g_L$. From the plot, we see that in order to give enough space for $M_{H_1}$, we need $x \gtrsim 1$.

\begin{figure}[!t]
  \centering
  \includegraphics[width=0.49\columnwidth]{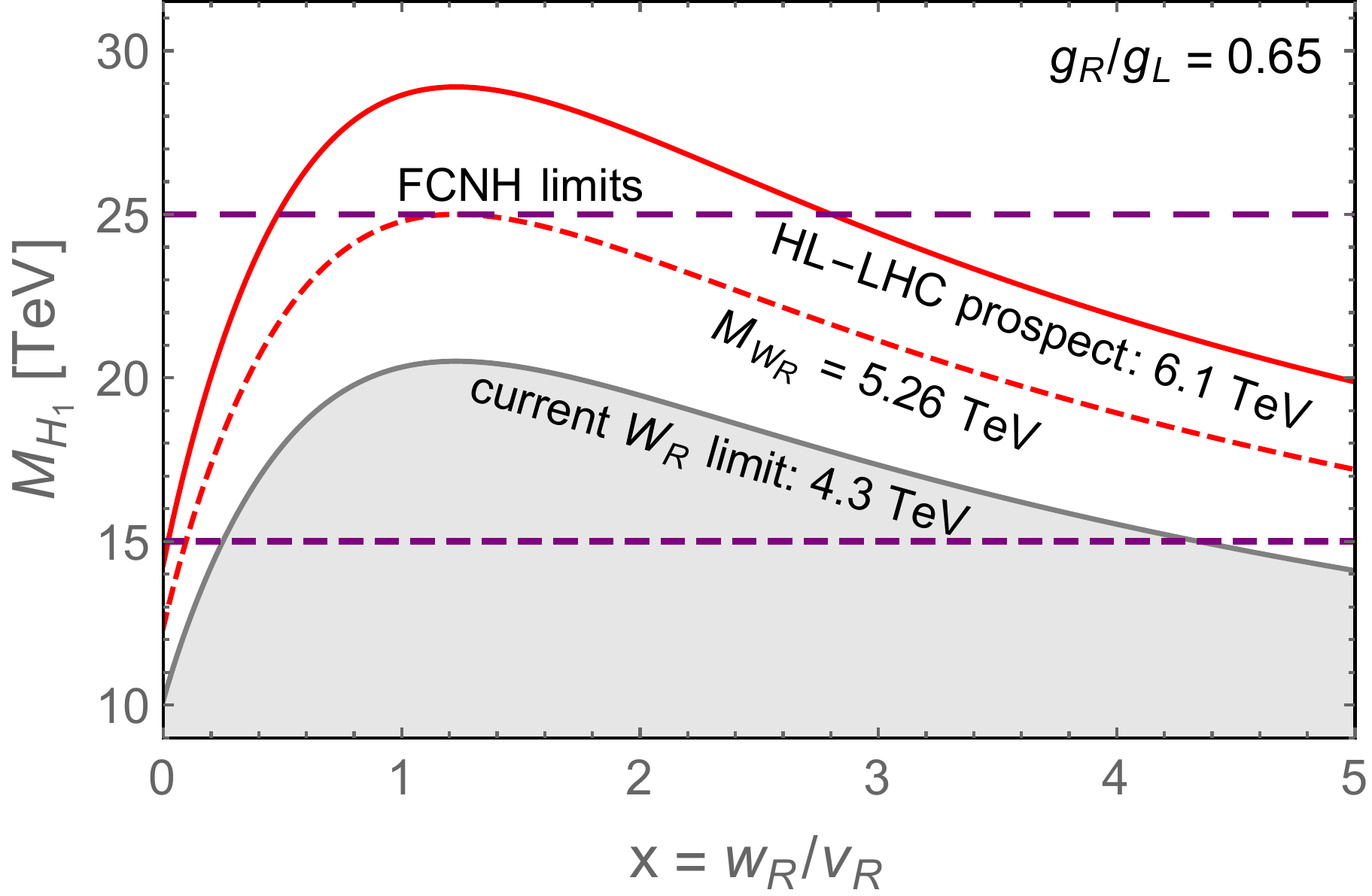}
  \includegraphics[width=0.49\columnwidth]{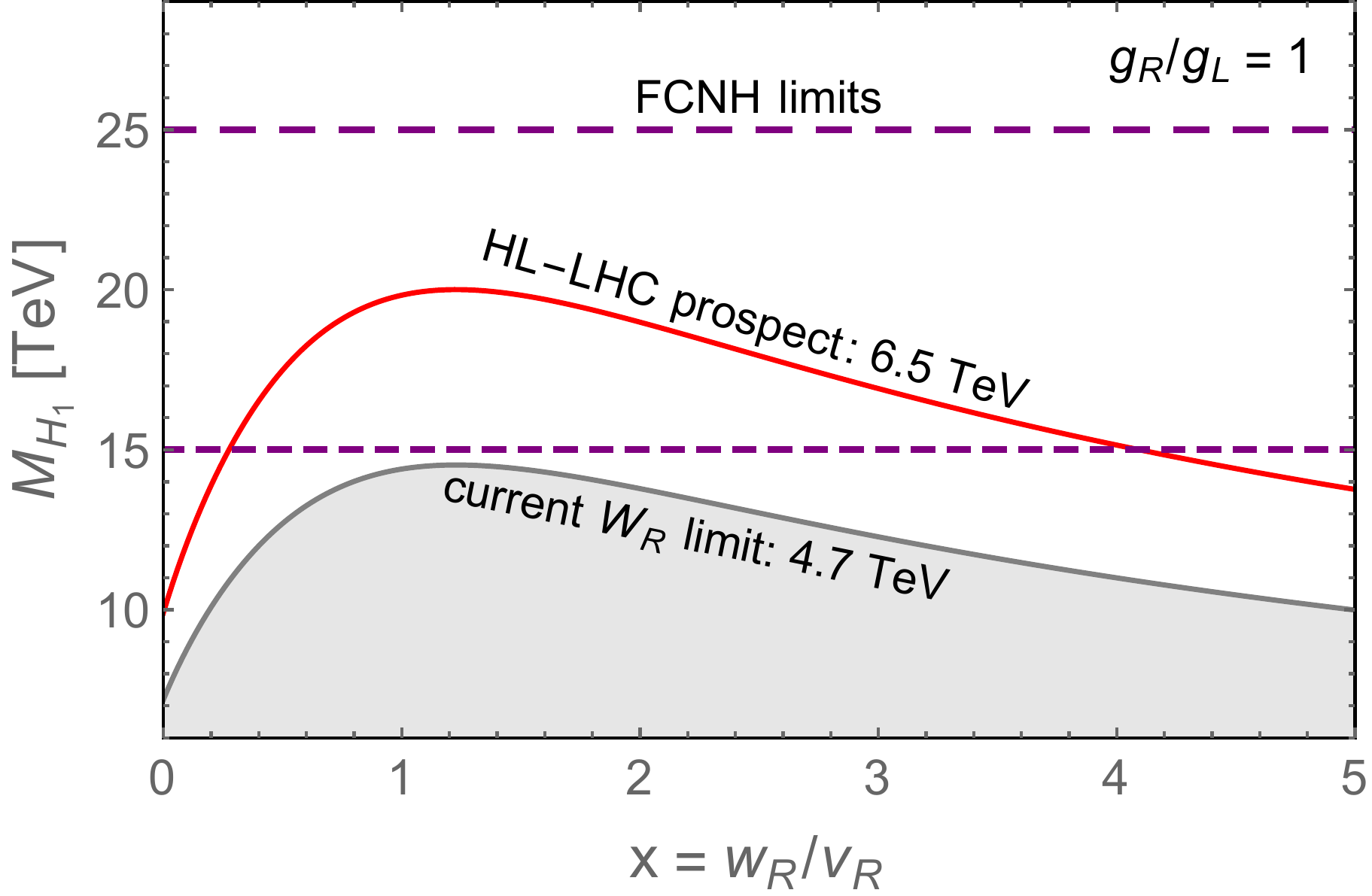}
  \caption{Heavy bidoublet scalar mass $M_{H_1}$ as function of the VEV ratio $x = w_R / v_R$, for the gauge coupling $g_R / g_L = 0.65$ (left) and 1 (right). The horizontal lines indicates the FCNH constraints $M_{H_1} > (15 - 25)$ TeV. The shaded regions are excluded by the current LHC constraints on $W_R$ boson mass, which are respectively 4.3 TeV and 4.7 TeV for $g_R / g_L = 0.65$ and 1~\cite{Chauhan:2018uuy}, while the solid red curve are the prospects at the HL-LHC, being respectively 6.1 TeV and 6.5 TeV. In the left panel the dashed red curve indicates the minimal value of $M_{W_R} = 5.26$ TeV for which we can have the scalar $H_1$ mass as high as 25 TeV. We have taken $\alpha_3 = \eta_1 = 1$ in Eq.~(\ref{eqn:MH1}) in the plots. }
  \label{fig:MH1b}
\end{figure}

In Fig.~\ref{fig:MH1b} we show explicitly the dependence of $M_{W_R}$ on the VEV ratio $x = w_R / v_R$ in the plane of $x - M_{H_1}$, for respectively the specific value of $g_R / g_L = 0.65$ in the left panel and $g_R/g_L = 1$ in the right panel. The gray shaded regions are excluded by the current LHC constraints on the $W_R$ mass, which are respectively 4.3 TeV and 4.7 TeV for the two benchmark values of $g_R$ (cf. Fig.~\ref{fig:WRreaches}). The prospects at the HL-LHC are respectively 6.1 TeV and 6.5 TeV, shown as the red curves~\cite{Chauhan:2018uuy}. As seen in the right panel of Fig.~\ref{fig:MH1b} and implied in the right panel of Fig.~\ref{fig:MH1}, for $g_R = g_L$ the FCNH limits on $H_1$ mass have precluded a large parameter space of $W_R$ reach at the HL-LHC, depending on the specific value of $H_1$ mass limit. For instance, with $M_{H_1} = 15$ TeV, the probable $x$ range at HL-LHC is $0.28 < x < 4.1$, while the mass $M_{H_1} = 25$ TeV is too high to leave any space for $x$.
When the gauge coupling $g_R$ is small, the VEVs $v_R$ and $w_R$ can be larger, and the viable parameter space becomes much larger: As shown in the left panel of Fig.~\ref{fig:MH1b}, if $M_{H_1} = 15$ TeV, the current $W_R$ limits require that $x < 0.25$ or $x > 4.3$, and a much larger range of $x > 0.02$ could be probed by the searches of $W_R$ boson at the HL-LHC. It should be noted the $W_R$ mass has to be larger than 3.16 TeV for the case of $M_{H_1} =15$ TeV, part of which region has been excluded by the current LHC data~\cite{Aaboud:2018spl, Sirunyan:2018pom}. Even if $M_{H_1} = 25$ TeV, $x$ can be probed in the range of $0.48 < x < 2.8$. We find that to have the scalar mass $M_{H_1}$ up to 25 TeV in the case of $g_R / g_L = 0.65$, the minimal $W_R$ mass is required to be 5.26 TeV, as indicated by the dashed red curve in the left panel of Fig.~\ref{fig:MH1b}.

The scalar mixing angle $\theta_3$ is also bounded by the FCNH data. In the limit of $M_3 \to 0$ and $\beta_1 \to 0$, plugging equation (\ref{eqn:MH1}) into equation (\ref{eqn:theta3}), we get
\begin{equation}
|\sin\theta_3| \ < \
\frac{1}{8\eta_1} \frac{v_{\rm EW}}{v_R}
\left(\frac{1}{x^3}+\frac{2\sqrt{2\pi} \eta_1^{1/4}}{x^2}\right) \,.
\end{equation}
A ${\cal O} (10 \, {\rm TeV})$ range $v_R$ implies an upper bound on the mixing angle $\theta_3$. For instance, if we set $\eta_1 = 1$ and $x = 1$, then a 10 TeV $v_R$ requires that $|\sin\theta_3| < 0.013$. In other words, if a sizable mixing of SM Higgs with other scalars could be found in future precision measurements, that might be in conflict with the model we are considering, and a large parameter space of our model will be excluded.


\subsection{The charged scalars}

In order to study the rich phenomenology of the model, we also need to know the approximate masses of the charged scalars of the model. It is straightforward to obtain the doubly-charged scalar $H^{\pm\pm}$, which is from the triplet $\Delta_R$, with the mass of $M_{H^{\pm\pm}}^2 = 4 \rho_2 v_R^2$. Regarding the singly-charged scalars, in the basis of $\{ \phi_1^{\pm} ,\, \phi_2^{\pm} ,\, \Delta_R^{\pm} ,\, \delta_R^{\pm} \}$, their mass matrix is given by
\begin{eqnarray}
\label{eqn:chargedscalarmatrix}
{\cal M}^\pm \ \simeq \
\left( \begin{matrix}
\tilde{\alpha}_3 \xi^2 & \tilde{\alpha}_3 \xi & \frac{\alpha_3 \varepsilon \xi}{\sqrt{2}} & -\frac{(\tilde{\alpha}_3 - \alpha_3)\varepsilon \xi}{\sqrt{2}x} \\
\tilde{\alpha}_3 \xi & \tilde{\alpha}_3 & \frac{\alpha_3 \varepsilon}{\sqrt{2}} & -\frac{(\tilde\alpha_3 - \alpha_3)\varepsilon}{\sqrt{2}x} \\
\frac{\alpha_3 \varepsilon \xi}{\sqrt{2}} & \frac{\alpha_3 \varepsilon}{\sqrt{2}} & \frac{rx}{\sqrt2} & \frac{r}{\sqrt2} \\
-\frac{(\tilde{\alpha}_3 - \alpha_3)\varepsilon \xi}{\sqrt{2}x} & -\frac{(\tilde\alpha_3 - \alpha_3)\varepsilon}{\sqrt{2}x} & \frac{r}{\sqrt2} & \frac{r}{\sqrt2 x}
\end{matrix} \right) v_R^2 \,.
\end{eqnarray}
This matrix can be diagonalized via, up to the first order of $\varepsilon$ and $\xi$,
\begin{eqnarray}
\label{eqn:chargedscalarmixing}
\left( \begin{array}{c}
G_W^{\pm} \\ H_1^{\pm} \\ G_{W_R}^{\pm} \\ H_2^{\pm} \\
\end{array} \right) \ \simeq \
\left(
\begin{array}{ccccc}
1 & -\xi & 0 & 0  \\
\xi  &  1 & \frac{\alpha_3 \varepsilon}{\sqrt{2} \tilde{\alpha}_3} & -\frac{(\tilde{\alpha}_3-\alpha_3) \varepsilon}{\sqrt{2}x\tilde{\alpha}_3} \\
 -\frac{\varepsilon}{\sqrt{2(1+x^2)}} & 0 & \frac{1}{\sqrt{1+x^2}} & -\frac{x}{\sqrt{1+x^2}}  \\
 0 & -\frac{[\tilde{\alpha}_3-\alpha_3(1+x^2)]\varepsilon}{\sqrt{2(1+x^2)}x\tilde{\alpha}_3} & \frac{x}{\sqrt{1+x^2}} & \frac{1}{\sqrt{1+x^2}} \\
\end{array} \right)
\left( \begin{array}{c}
\phi_1^{\pm} \\ \phi_2^{\pm} \\ \Delta_R^{\pm} \\ \delta_R^{\pm}  \\
\end{array} \right) \,.
\end{eqnarray}
There are two massless eigenstates of this matrix: $G_{W}$ and $G_{W_R}$, which become the longitudinal modes of the $W$ and $W_R$ gauge bosons, leaving only the two massive singly-charged scalars $H_1^\pm$ and $H_2^\pm$. As in the minimal LRSM, at leading order the mass of $H_1^\pm$ is degenerate with that of $H_1$ and $A_1$. The scalar $H_2^\pm$ is new in the extended model, with mass
\begin{equation}
\label{eqn:mass3}
M_\pm^2 \ \simeq \
\frac{r (1+x^2)}{\sqrt2 x} v_R^2 \,.
\end{equation}
As we will see in the following sections, the presence of $H_2^\pm$ induces very rich phenomenologies in the LRSM, including the decay of $W_R$ boson and the heavy RHNs.

\section{The singly-charged scalar $H_2^\pm$}
\label{sec:singlychargedscalar}

\subsection{$H_2^\pm$ Decay}
\label{sec:H2decay}

The singly-charged scalar $H_2^\pm$ is new beyond the minimal LRSM. As shown in Eqs.~(\ref{eqn:chargedscalarmatrix}) and (\ref{eqn:chargedscalarmixing}), at leading order it is a linear combination of $\Delta_R^\pm$ and $\delta_R^\pm$ with the mixing angle $\tan\varphi_\pm = x$, with a subdominant portion from mixing with the heavy singly-charged scalar $H_1^\pm \simeq \phi_2^\pm$, cf. Eq.~(\ref{eqn:chargedscalarmixing}). One should note that if $x \sim {\cal O} (1)$, the mixing of $H_2^\pm$ with $H_1^\pm$ is approximately $-\frac{1}{2}\varepsilon$ and does not depend on any quartic coupling or $\tilde{\alpha}_3$. $H_1^\pm$ couples to the SM quarks~\cite{Zhang:2007da}, therefore $H_2^\pm$ decays predominately into the SM quarks via the mixing with $H_1^\pm$:
\begin{eqnarray}
H_2^\pm \to q \bar{q}
\end{eqnarray}

To accommodate the type-I seesaw and generate the RHN masses, the triplet $\Delta_R$ couples to the right-handed lepton doublets $\psi_R = (\ell, \, N)^{\sf T}$ via the Yukawa Lagrangian
\begin{eqnarray}
\label{eqn:Yukawa}
{\cal L}_{\rm Yukawa} \ = \
f_R \psi_R^{\sf T} C i\sigma_2 \Delta_R \psi_R \,.
\end{eqnarray}
From this Lagrangian, the singly-charged scalar $H_2^\pm$ couples to the charged leptons and RHNs, then we have the decay channel, if kinematically allowed,
\begin{eqnarray}
H_2^\pm \to \ell_\alpha^\pm N_\beta \,,
\end{eqnarray}
with $\alpha,\, \beta = e,\, \mu,\, \tau$ the flavor indices. The lepton flavors indices $\alpha$ and $\beta$ might be the same or different, depending the Yukawa coupling matrix $(f_R)_{\alpha\beta}$ and the RHN mixing. The partial widths $\Gamma (H_2^\pm \to q \bar{q})$ and $\Gamma (H_2^\pm \to \ell_\alpha^\pm N_\beta)$ are collected in Appendix~\ref{sec:decay}.

Neglecting the small heavy-light neutrino mixing which is strongly constrained~\cite{Deppisch:2015qwa}, the RHNs could decays via the gauge interactions with the $W_R$ boson, i.e. $N_\alpha \to \ell_\beta^\pm W_R^{\mp (\ast)} \to \ell_\beta^\pm q\bar{q}$. Then, as a result of the Majorana nature of the RHNs, the final states of the decay $H_2^\pm \to \ell_\alpha^\pm N_\beta$ consist of the opposite-sign and same-sign dileptons plus jets
\begin{eqnarray}
H_2^\pm \ \to \ \ell_\alpha^+ \ell_\beta^- q \bar{q} \,, \quad
\ell_\alpha^\pm \ell_\beta^\pm q \bar{q} \,.
\end{eqnarray}
which are quite similar to the decay of $W_R$ boson in the LRSM.\footnote{In principle we have also the decays $H_2^\pm \to \ell_\alpha^{+ (\pm)} \ell_\beta^{- (\pm)} W_R^{\pm (\mp) (\ast)}$, with the subsequent decays $W_R^{(\ast)} \to WZ,\, Wh,\, W H_3^{(\ast)}$, however, these channels are all highly suppressed by the small ratios $\xi^2 = (\kappa'/\kappa)^2$ or $\varepsilon^2 = (v_{\rm EW} / v_R)^2$, and thus disregarded here.} As for the $W_R$ boson, the same-sign dilepton channel $H_2^\pm \to \ell_\alpha^\pm \ell_\beta^\pm jj$ is the most promising channel to search for the singly-charged scalar $H_2^\pm$ at hadron and lepton colliders, which is almost background free.




\subsection{Production at hadron and lepton colliders}
\label{sec:production}

The decay of singly-charged scalar $H_2^\pm \to \ell_\alpha^\pm \ell_\beta^\pm q \bar{q}$ is reminiscent of the $W_R$ boson, with the scalar and vector bosons sharing the same signal of same-sign dilepton plus quark jets. If such ``smoking-gun'' signatures can be found at LHC or future higher luminosity/energy colliders, it is {\it not} necessarily from the $W_R$ gauge boson. We would like to emphasize here that though the decay products are the same, the scalar $H_2^\pm$ is actually very different from the $W_R$ boson, 
because its spin is different from the $W_R$ boson.

Just like the neutral scalar $H_3$ in the minimal LRSM, $H_2^\pm$ also does not couple directly to the quark sector and is thus hadrophobic before the electroweak symmetry breaking. The couplings to the SM quarks come mainly from the small mixing of $\delta_R^\pm$ with the component $\phi_2^\pm$ from the bidoublet.
As a consequence, the production of $H_2^\pm$ in hadron and lepton colliders are very different from the $W_R$ boson which couples directly to the SM quarks.

Through the gauge interactions with the $\gamma$, $Z$ and $Z_R$ bosons, $H_2^\pm$ can be produced via the process $pp,\, e^+ e^- \to \gamma^\ast/Z^\ast/Z_R^{(\ast)} \to H_2^+ H_2^-$. Furthermore, if $M_{H_2^\pm} + M_N < M_{W_R}$, it could be produced directly from $W_R$ boson decay in hadron colliders via $pp \to W_R^{\pm (\ast)} \to H_3 H_2^\pm$. This is a new decay channel of the $W_R$ boson beyond the minimal LRSM. More details about this channel can be found in Section~\ref{sec:WR}. In future lepton colliders, high luminosity photon beams can be obtained by Compton backscattering of low energy high intensity laser beam off the high energy electron beam~\cite{Ginzburg:1981vm, Ginzburg:1982yr, Telnov:1989sd}, and the singly-charged scalar $H_2^\pm$ can also be produced through the process $e^\pm\gamma \to e^{\pm\ast} \to N H_2^\pm$.

\begin{figure}[!t]
  \centering
  \includegraphics[height=0.18\textwidth]{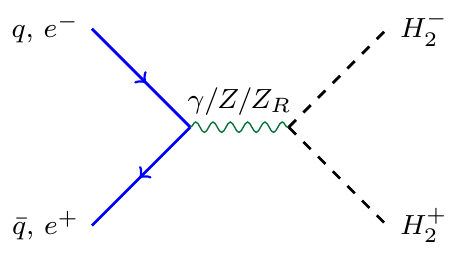}
  \includegraphics[height=0.18\textwidth]{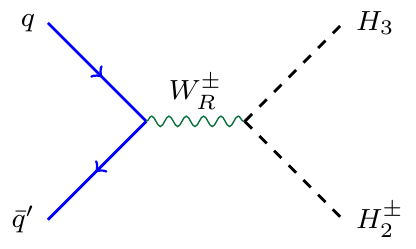}
  \includegraphics[height=0.18\textwidth]{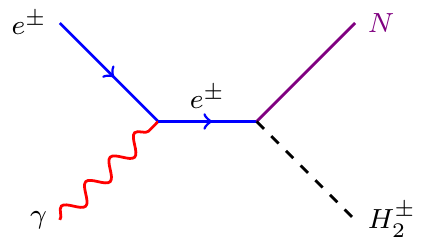}
  \caption{Feynman diagrams for the production at hadron and lepton colliders via the process $q\bar{q},\, e^+ e^- \to \gamma/Z/Z_R \to H_2^+ H_2^-$ (left), $q \bar{q} \to W_R^\pm \to H_3 H_2^\pm$ (middle) and $e^\pm \gamma \to N H_2^\pm$ (right).}
  \label{fig:diagram_production}
\end{figure}

All these channels are collected in Table~\ref{tab:production} and the corresponding Feynman diagrams are presented in Fig.~\ref{fig:diagram_production}. 
Combining the production and subsequent decays of $H_3 \to \gamma\gamma$~\cite{Dev:2016vle, Dev:2017dui} and  $H_2^\pm \to \ell^\pm N \to \ell^\pm \ell^\pm jj$, the final states for these channels are also shown in Table~\ref{tab:production}. The particles in parenthesis are from $H_3$ or $H_2^{\pm\pm}$ decay; the reconstructed invariant mass of these particles are expected to be close to the mother scalar particle masses.

\begin{table}[!t]
  \centering
  \caption{Production channels of $H_2^\pm$ at hadron and lepton colliders and the corresponding final states, assuming the most promising decay modes $H_3 \to \gamma\gamma$~\cite{Dev:2016vle, Dev:2017dui} and $H_2^\pm \to \ell_\alpha^\pm \ell_\beta^\pm jj$. The subscripts $\alpha,\, \beta,\,\gamma,\,\delta,\,\eta$ are the lepton flavor indices.}
  \label{tab:production}
  \begin{tabular}{ll}
  \hline\hline
  channel  & final states \\ \hline
  $pp,\, e^+ e^- \to \gamma^\ast/Z^\ast/Z_R^{(\ast)} \to H_2^+ H_2^-$ &
  $(\ell_\alpha^+ \ell_\beta^+ jj) (\ell_\gamma^- \ell_\delta^- jj)$ \\ \hline
  $pp \to W_R^{\pm (\ast)} \to H_3 H_2^\pm$ &
  $(\gamma\gamma) (\ell_\alpha^\pm \ell_\beta^\pm jj)$ \\ \hline
  $e^\pm\gamma \to e^{\pm\ast} \to N H_2^\pm$ &
  $\ell_\alpha^\pm (\ell_\beta^\mp \ell_\gamma^\mp jj)  (\ell_\delta^\pm \ell_\eta^\pm jj)$ \\ \hline\hline
  \end{tabular}
\end{table}

To proceed to estimate the production cross sections at future hadron and lepton colliders, we first check the limits on the VEV $v_R$ in the extended LRSM from the direct searches of $W_R$ and $Z_R$ bosons at LHC 13 TeV and Eq.~(\ref{eqn:MH1}). In the minimal LRSM without the triplet $\delta_R$, the $W_R$ and $Z_R$ boson masses are respectively $M_{W_R}^2 = g_R^2 v_R^2$ and $M_{Z_R}^2 = 2 (g_R^2 + g_{BL}^2) v_R^2$ with the predictive relation $M_{Z_R} \simeq 1.7 M_{W_R}$. When the LRSM is extended in the scalar sector, like the case in this paper, the mass relation between the $W_R$ and $Z_R$ bosons does not hold true any more, and the $Z_R$ boson could even be lighter than the $W_R$ boson~\cite{Patra:2015bga}. With the benchmark values of $g_R = g_L$ and $x = w_R/v_R = 1$, the direct searches of $W_R$ boson imply that  $v_R \gtrsim 5.9$ TeV, as seen in the right panel of Fig.~\ref{fig:MH1}. The most stringent constraints on the $Z_R$ are from the dilepton data at LHC in the channel $pp \to Z_R \to \ell^+ \ell^-$ (with $\ell = e,\,\mu$). Following~\cite{Chauhan:2018uuy}, we rescale the production cross section $\sigma (pp \to Z_R \to \ell^+ \ell^-)$ with respect to that for a sequential $Z'$ boson, and the latest ATLAS and CMS data~\cite{ATLAS:2016cyf,CMS:2016abv} requires that $M_{Z_R} > 3.7$ TeV which implies that $v_R > 3.4$ TeV when we set $g_R = g_L$. In the extended LRSM we are considering, the FCNH limits $M_{H_1} > 15$ TeV impose a lower bound on the $v_R$ scale at loop level, as seen in Eq.~(\ref{eqn:MH1}). Following Section~\ref{sec:loop}, setting $\alpha_3 = \eta = 1$ leads to $v_R \gtrsim 6$ TeV, as seen in the right panel of Fig.~\ref{fig:MH1}.

\begin{figure}[!t]
  \centering
  \includegraphics[width=0.495\columnwidth]{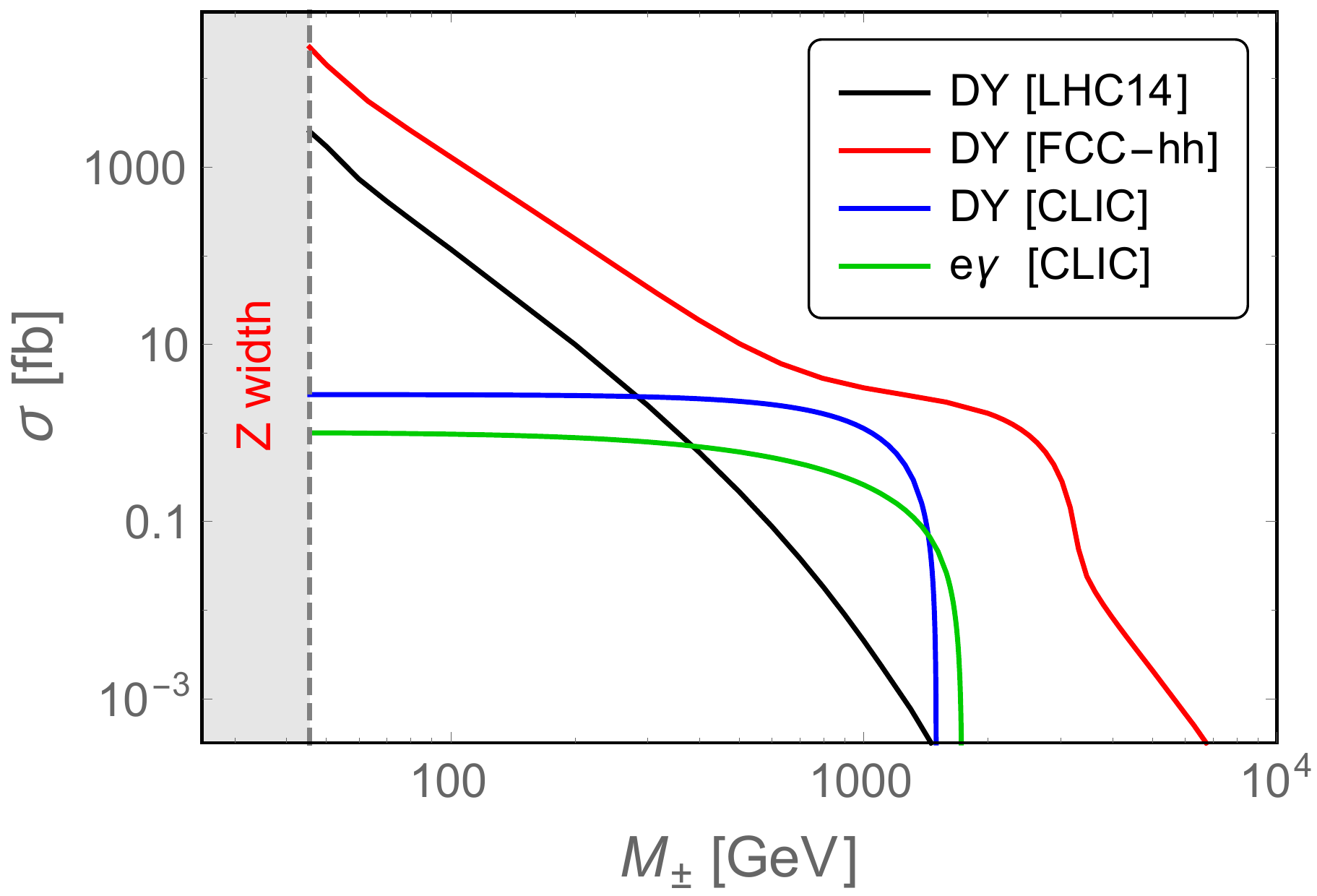}
  \includegraphics[width=0.49\columnwidth]{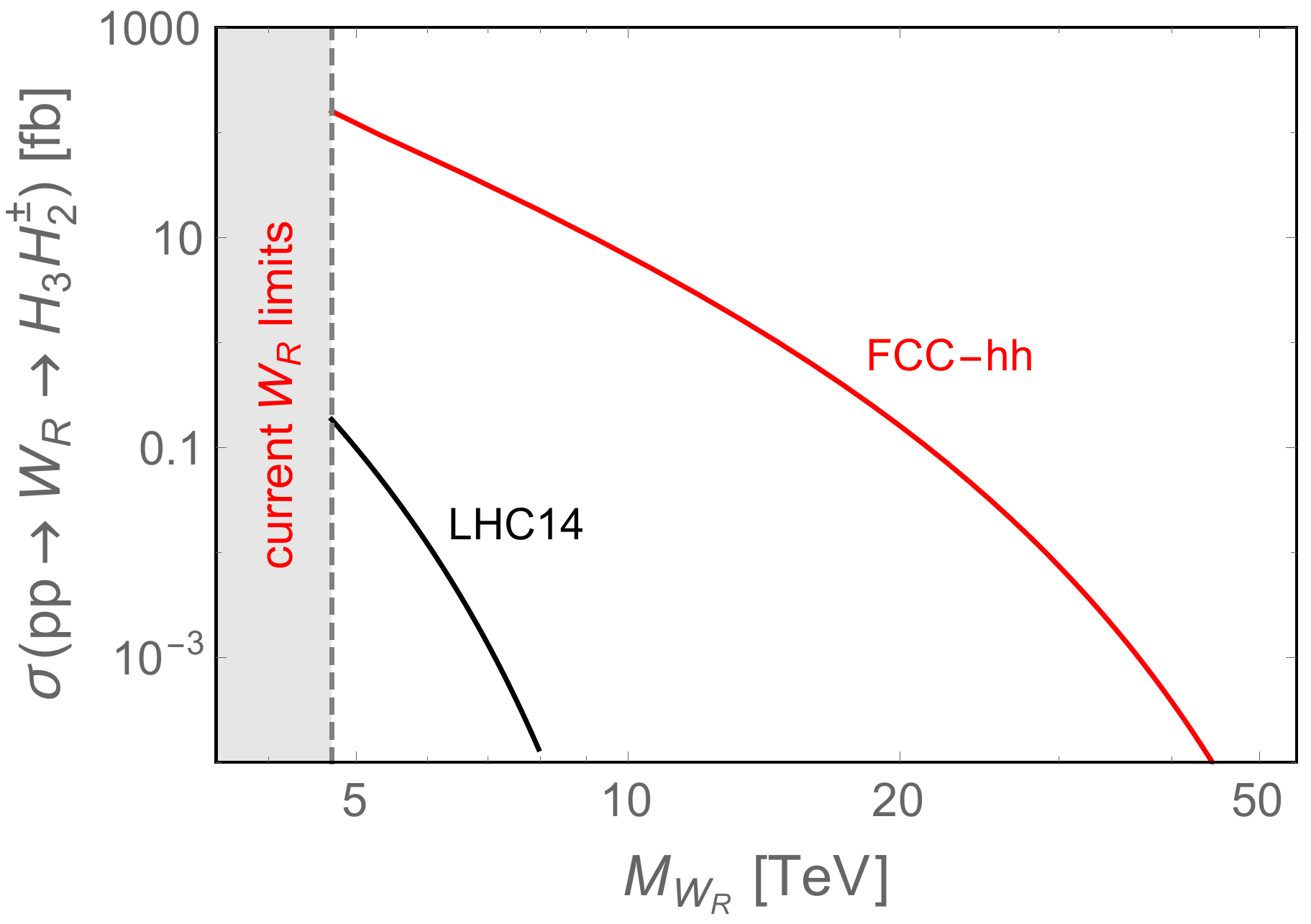}
  \caption{Production cross sections for $H_2^\pm$ at hadron and lepton colliders: The contours in the left panel are for he Drell-Yan channel $pp,\, e^+ e^- \to H_2^+ H_2^-$ at LHC 14 TeV, the 100 TeV collider FCC-hh, CLIC 3 TeV and the process $e^\pm \gamma \to N H_2^\pm$ at CLIC 3 TeV, as function of $H_2^\pm$ mass. The gray band is excluded by the precision measurement of the $Z$ boson width.
  The right panel is for the channel $pp \to W_R \to H_3 H_2^\pm$ at LHC 14 TeV and FCC-hh as function of $W_R$ mass, with the gray band excluded by current LHC constraints on $W_R$ boson mass~\cite{Aaboud:2018spl, Sirunyan:2018pom}. We have taken $v_R = 6$ TeV for the left panel, $g_R = g_L$, $x = w_R/v_R = 1$, the RHN mass $M_N = 1$ TeV and $\sin\theta_5 = 0$. The Feynman diagrams are shown in Fig~\ref{fig:diagram_production}. See text for more details. }
  \label{fig:production}
\end{figure}

Respecting all the constraints on $v_R$, we set explicitly $v_R = 6$ TeV as well as $g_R = g_L$ and $x = w_R/v_R = 1$. Then the $Z_R$ boson mass $M_{Z_R} = 6.6$ TeV. The production cross sections for the Drell-Yan process $pp \to H_2^+ H_2^-$ at LHC 14 TeV and the future 100 TeV collider FCC-hh~\cite{FCC-hh} and $e^+ e^- \to H_2^+ H_2^-$ at CLIC 3 TeV~\cite{Battaglia:2004mw} are shown in the left panel of Fig.~\ref{fig:production}, depicted respectively in black, red and blue. In this plot We have included a $k$-factor of $1.18$ for the production of $H_2^\pm$ at hadron colliders to account for the high-order corrections~\cite{Hamberg:1990np}.
The 6.6 TeV $Z_R$ boson is too heavy to be produced directly at LHC and CLIC and thus could hardly play any role for the Drell-Yan process. However, it is important for the production of $H_2^\pm$ at FCC-hh, and the bump-like structure in the left panel of Fig.~\ref{fig:production} are due to the $Z_R$ resonance~\cite{Dev:2016dja}. The gray band in the left panel of Fig.~\ref{fig:production} indicates the constraint on $H_2^\pm$ mass from precision measurements of SM $Z$ boson width~\cite{Tanabashi:2018oca}, i.e. $M_\pm > m_Z/2$, as the singly-charged scalar contributes to $Z$ boson decay via the channel $Z \to H_2^+ H_2^-$ when it is lighter than $m_Z/2$.

The cross section for associated production of $H_2^\pm$ with a RHN in the channel $e^\pm \gamma \to N_e H_2^\pm$ at CLIC 3 TeV is also shown in the left panel of Fig.~\ref{fig:production}, in the green color. This process is induced by the Yukawa couplings $(f_R)_{e\beta}$ in Eq.~(\ref{eqn:Yukawa}), which induces lepton flavor violation if $\beta \neq e$. We are not focusing on the flavor structure in this paper; for the sake of concreteness, we consider only the electron-flavored RHN $N_e$ in the associated production and assume the mixing of $N_e$ with  the other two RHNs $N_{\mu,\,\tau}$ are small. As a consequence, the Yukawa coupling $(f_R)_{ee}$ is related to the RHN mass $M_N$ via $(f_R)_{ee} = M_N / 2 v_R$.

The singly-charged scalar $H_2^\pm$ could also produced at hadron colliders from $W_R$ decay, as shown in Fig.~\ref{fig:diagram_production} and Table~\ref{tab:production}.  The production cross sections times branching ratio $\sigma (pp \to W_R \to H_3 H_2^\pm)$ at LHC 14 TeV and FCC-hh are shown in the right panel of Fig.~\ref{fig:production}, as function of $M_{W_R}$. The branching fraction ${\rm BR} (W_R \to H_3 H_2^\pm)$ depends largely on the factor of mixing angles $\sin^2(\theta_5 + \varphi_\pm)$ in Eq.~(\ref{eqn:WRdecay7}). $x = 1$ implies that $\varphi = \arctan{x} = 45^\circ$. To be concrete, we assume $\sin\theta_5 \simeq 0$ (thus $\sin^2(\theta_5 + \varphi_\pm) = 1/2$), $H_3$ being light (say $M_{H_3} \lesssim 100$ GeV)~\cite{Dev:2016vle, Dev:2017dui}, and $H_2^\pm$ being significantly lighter than the $W_R$ boson in the right panel of Fig.~\ref{fig:production}. The gray band in this plot is excluded by current LHC constraints of 4.7 TeV on $W_R$ boson mass for the special case of $g_R = g_L$~\cite{Aaboud:2018spl, Sirunyan:2018pom}.

It is clear in Fig.~\ref{fig:production} that the singly-charged scalar $H_2^\pm$ can be produced abundantly at both hadron and lepton colliders in a large region of parameter space. Given the benchmark values of parameters we have chosen, the production cross sections could reach the order of ${\cal O} (10^{-3}\, {\rm fb})$ for $M_\pm \lesssim 1$ TeV in the Drell-Yan and $e\gamma$ processes. For the production of $H_2^\pm$ from $W_R$ decay, the production cross section times branching ratio $\sigma (pp \to W_R \to H_3 H_2^\pm)$ is larger than $10^{-3}$ fb at LHC 14 TeV (FCC-hh) if $M_{W_R} < 7.1 \, (37)$ TeV.


%


\section{RHN Decay}
\label{sec:RHN}

The Yukawa Lagrangian in Eq.~(\ref{eqn:Yukawa}) has also profound implications for the decay of heavy RHNs in the LRSM. In addition to the ``standard'' decay through the gauge couplings to the $W_R$ boson $N_\alpha \to \ell^\pm W_R^{\mp \, (\ast)} \to \ell_\beta^\pm q\bar{q}$, the RHNs have a new decay channel through the Yukawa couplings in the extended LRSM, i.e. $N_\alpha \to \ell_\beta^\pm H_2^{\mp}$, if kinematically allowed.  With the mass relation $M_{\pm} < M_N < M_{W_R}$, it is expected that the two-body decays $N_\alpha \to \ell_\beta^\pm H_2^{\mp}$ dominate over the three-body channels $N_\alpha \to \ell_\beta^\pm q\bar{q}$ via the $W_R$ boson, and the two jets from the decay $H_2^\pm \to q \bar{q}$ form a peak in vicinity of the $H_2^\pm$ mass, which could be easily distinguished from the continuum spectrum of the invariance mass $m_{qq}$ from the three-body decays. Even if $M_N < M_{\pm}$, the scalar $H_2^\pm$ contributes also to the three-body decay $N_\alpha \to \ell_\beta^\pm q \bar{q}$, as in this case we have both the $W_R$ and $H_2^\pm$ propagators for the three-body decay. This is very important when the RHN $N_\alpha$ has a mass in the range $\sim (1 - 100)$ GeV and form displaced vertices in the high energy and high intensity experiments~\cite{Helo:2013esa, Castillo-Felisola:2015bha, Cottin:2019drg}. This is, however, beyond the main goal of this paper and is postponed to a future publication.

\section{$W_R$ Decay}
\label{sec:WR}


In the minimal LRSM, the $W_R$ boson decays predominately into the SM quarks as well as charged leptons plus heavy neutrinos $N$ if kinematically allowed, i.e. $W_R \to q \bar{q},\, \ell N$. From the scalar and $W - W_R$ mixing, $W_R$ decays also into the light $W$, $Z$ bosons and the scalars, e.g. $W_R \to WZ,\, Wh,\, WH_3$. However, these bosonic channels are suppressed either by $\xi = \kappa'/\kappa$ or by $\varepsilon = v_{\rm EW}/v_R$. In presence of the extra triplet $\delta_R$ and the neutral and charged scalar mixings beyond the minimal LRSM, new decay channels open for the heavy $W_R$ boson. In particular, the scalars $H_3$ and $H_2^\pm$ are both from the right-handed sector, which couple directly to the $W_R$ boson and induce the new decay channel $W_R \to H_2^\pm H_3$. The width for this channel is proportional to the combination of the neutral and charged scalar mixing angles $\sin (\theta_5 + \varphi_\pm)$ and is not suppressed by any small parameters. There is also the decay channel $W_R \to H_2^\pm h$, which is however suppressed by the small parameter $\varepsilon^2$. All the partial widths for these decay channels are collected in Appendix~\ref{sec:decay}. In the limit of $M_{W_R} \gg M_{N,\,H_3,\,H_2^\pm}$, the branching fractions of the unsuppressed decays $W_R \to q\bar{q},\, \ell N,\, H_2^\pm H_3$ depend only on the degrees of freedom and the mixing angle $\sin (\theta_5 + \varphi_\pm)$:
\begin{eqnarray}
\Gamma (W_R \to q\bar{q}):
\Gamma (W_R \to \ell N):
\Gamma (W_R \to H_2^\pm H_3) \ \simeq \
9 : 3 : \sin^2 (\theta_5 + \varphi_\pm) \,.
\end{eqnarray}
For instance, in the limit of $\theta_5 \to 0$ and $\tan\varphi_\pm = x = 1$ we have $\sin^2 (\theta_5 + \varphi_\pm) = 1/2$ and the branching fraction
\begin{eqnarray}
{\rm BR} (W_R \to q\bar{q}) \ \simeq \ \frac{18}{25} \,, \quad
{\rm BR} (W_R \to \ell N) \ \simeq \ \frac{6}{25} \,, \quad
{\rm BR} (W_R \to H_2^\pm H_3) \ \simeq \ \frac{1}{25} \,.
\end{eqnarray}
As a result of the new decay channel $W_R \to H_2^\pm H_3$, the LHC constraint on the $W_R$ boson mass from the searches of same-sign dilepton plus jets is slightly weaker than in the minimal LRSM: in the extended LRSM the branching fraction ${\rm BR} (W_R \to \ell^\pm \ell^\pm jj)$ has to be rescaled with respect to that in the minimal LRSM by a factor of $12 / (12 + \sin^2 (\theta_5 + \varphi_\pm))$, which is $\simeq 24/25$ for the special case of ${\rm BR} (W_R \to H_2^\pm H_3) \simeq 1/25$.

In the minimal LRSM, the primary production channel for the hadrophobic neutral scalar $H_3$ is through associated production with the $W_R$ boson, i.e. $pp \to W_R^{\ast} \to W_R H_3$~\cite{Dev:2016vle, Dev:2017dui}. When the decays $H_3 \to \gamma\gamma$ and $W_R \to \ell^\pm \ell^\pm jj$ are taken into account, the final states for $pp \to H_3 W_R \to (\gamma\gamma) (\ell^\pm \ell^\pm jj)$ are the same as that in the process $pp \to W_R \to H_3 H_2^\pm \to (\gamma\gamma) (\ell^\pm \ell^\pm jj)$ in the extended LRSM, as shown in Table~\ref{tab:production}. However, as the experimental constraints on the singly-charged scalar $H_2^\pm$ is much weaker than that for the $W_R$ boson, $H_2^\pm$ could be much lighter than the $W_R$ boson and the production cross section $\sigma (pp \to W_R \to H_3 H_2^\pm)$ could reach a higher value than that for the $pp \to W_R H_3$ process~\cite{Dev:2016vle, Dev:2017dui}.

\section{Effects on DM in the LRSM}
\label{sec:DM}

In the minimal LRSM or in the current extended extension, there is no suitable candidate which can play the role of DM in the universe, for which there seems to be overwhelming evidence. In order to make the model more encompassing of observations, in recent years an extension has been suggested by adding for example a $B-L=0$ right-handed triplet fermion $\Psi$~\cite{Heeck:2015qra, Mambrini:2015vna, Garcia-Cely:2015quu, Berlin:2016eem, Patra:2015qny, Borah:2017xgm}. The neutral components of the triplet $\Psi$ is naturally stable without introducing any extra symmetry and is the DM candidate.  There is however one issue of the model~\cite{Garcia-Cely:2015quu}: At the tree level, all three members of $\Psi$ are degenerate; their masses get split only at the one-loop level and typically the mass splitting is in the few GeV range for few-TeV scale $W_R$ boson. Therefore, in the early universe, the freezing-out of DM is dominated by the co-annihilation processes like $\psi^0 \psi^\pm \to W_R \to f \bar{f}$, with $\psi^0$ and $\psi^\pm$ respectively the neutral (DM) and charged components from $\Psi$, and $f$ the SM fermions. However, as a result of the severe LHC constraints on $W_R$ boson mass~\cite{Aaboud:2018spl, Sirunyan:2018pom} and the indirect limits from gamma-ray flux measurements by H.E.S.S.~\cite{Abramowski:2011hc, Abramowski:2013ax, Abdalla:2016olq}, the simple triplet DM model has been almost excluded in the minimal LRSM~\cite{Heeck:2015qra, Garcia-Cely:2015quu}.

\begin{figure}[!t]
  \centering
  \includegraphics[width=0.3\columnwidth]{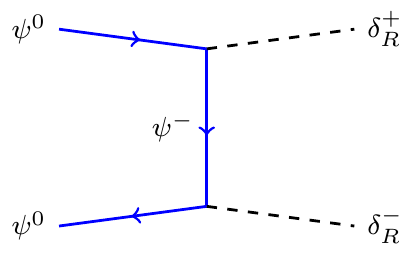}
  \includegraphics[width=0.3\columnwidth]{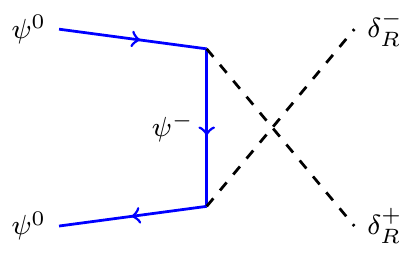}
  \caption{Feynman diagrams for DM annihilation $\psi^0 \psi^0 \to \delta_R^+ \delta_R^-$ in the extended LRSM.}
  \label{fig:diagram:DM}
\end{figure}

In the extended LRSM we are considering, however, due to the presence of the $\delta_R$ field, the DM phenomenology is very different, and we can find large parameter space to accommodate the DM particle from $\Psi$. In particular, there is a new coupling to right-handed fermion triplet of the form
\begin{eqnarray}
\label{eqn:Yukawa:DM}
{\cal L}_{\rm DM} \ \supset \
h_\psi \Psi^T C^{-1} \delta_R \Psi \,.
\end{eqnarray}
After symmetry breaking this coupling  splits the charged member of the triplet from the neutral one at the tree level:
\begin{eqnarray}
m_{\pm} \ = \ m_{\rm DM} + h_\psi w_R \,,
\end{eqnarray}
where $m_{\rm DM}$ and $m_{\pm}$ are respectively the mass for $\psi^0$ and $\psi^\pm$.
As long as the Yukawa coupling $h_\psi$ is not very small, there is no tree-level mass degeneracy in the $\Psi$ components, and the DM particles annihilate mainly in the channel
\begin{eqnarray}
\psi^0 \psi^0 \to \delta_R^+ \delta_R^- \,,
\end{eqnarray}
which is mediated by the charged component $\psi^\pm$ in both the $t$- and $u$-channels (if written in the mass eigenstates, here $\delta_R^\pm$ refers to the physical scalar $H_2^\pm$ with a mixing angle, which we absorb in $h_\psi$). The Feynman diagrams are shown in Fig.~\ref{fig:diagram:DM}. The DM annihilation cross section $\sigma (\psi^0 \psi^0 \to \delta_R^+ \delta_R^-)$ is given in Appendix~\ref{sec:DM:cs}, with the thermal averaged cross section times velocity
\begin{eqnarray}
\langle \sigma v \rangle \ = \
\frac{g_{\rm DM}^2 m_{\rm DM}}{64\pi^4 x n_{\rm eq}^2}
\int_{4 m_{\rm DM}^2}^{\infty} {\rm d}s \, \hat{\sigma}(s) \sqrt{s}
K_1 \left( \frac{x\sqrt{s}}{m_{\rm DM}} \right) \,,
\end{eqnarray}
where $g_{\rm DM} = 2$ is the number of degrees of freedom of DM $\psi^0$, $n_{\rm eq} = s (m_{\rm DM}) Y_{\rm eq}/x^3$ is the DM number density, $s (m_{\rm DM}) = \frac{2\pi^2}{45} g_\ast m_{\rm DM}^3$ is the entropy density ($g_\ast \simeq 110$ is the degrees of freedom for temperature in the TeV range) and $Y_{\rm eq} = \frac{g_{\rm DM}}{2\pi^2} \frac{x^2 m_{\rm DM}^3}{s(m_{\rm DM})} K_2(x)$. Here $K_{1,2}$ are respectively the modified Bessel function of the first and second kind, and $\hat{\sigma} (s) = 2 (s-4m_{\rm DM}^2) \sigma (s)$ is the reduced cross section. Considering the so-called instantaneous freeze-out approximation for solving the Boltzmann equation, the relic density of DM is
\begin{eqnarray}
\Omega_{\rm DM} h^2 \ = \
\frac{1.03 \times 10^9 \, {\rm GeV}^{-1}}{\sqrt{g_\ast} M_{\rm Pl}}
\left( \int_{x_f}^{\infty} \frac{\langle \sigma v \rangle}{x^2} {\rm d}x \right)^{-1} \,,
\end{eqnarray}
where $M_{\rm Pl} = 1.22 \times 10^{19}$ GeV is the Planck scale,  $x_f = m_{\rm DM}/T_f$ with $T_f$ the freeze-out temperature. $x_f$ can obtained by calculating the temperature at which the DM annihilation rate drops below the Hubble expansion rate
\begin{eqnarray}
x_f \ = \ \log
\left( \frac{0.038 \, g_{\rm DM} x_f^{1/2} m_{\rm DM} M_{\rm Pl} \langle \sigma v \rangle}{\sqrt{g_\ast}} \right) \,.
\end{eqnarray}

\begin{figure}[!t]
  \centering
  \includegraphics[width=0.495\columnwidth]{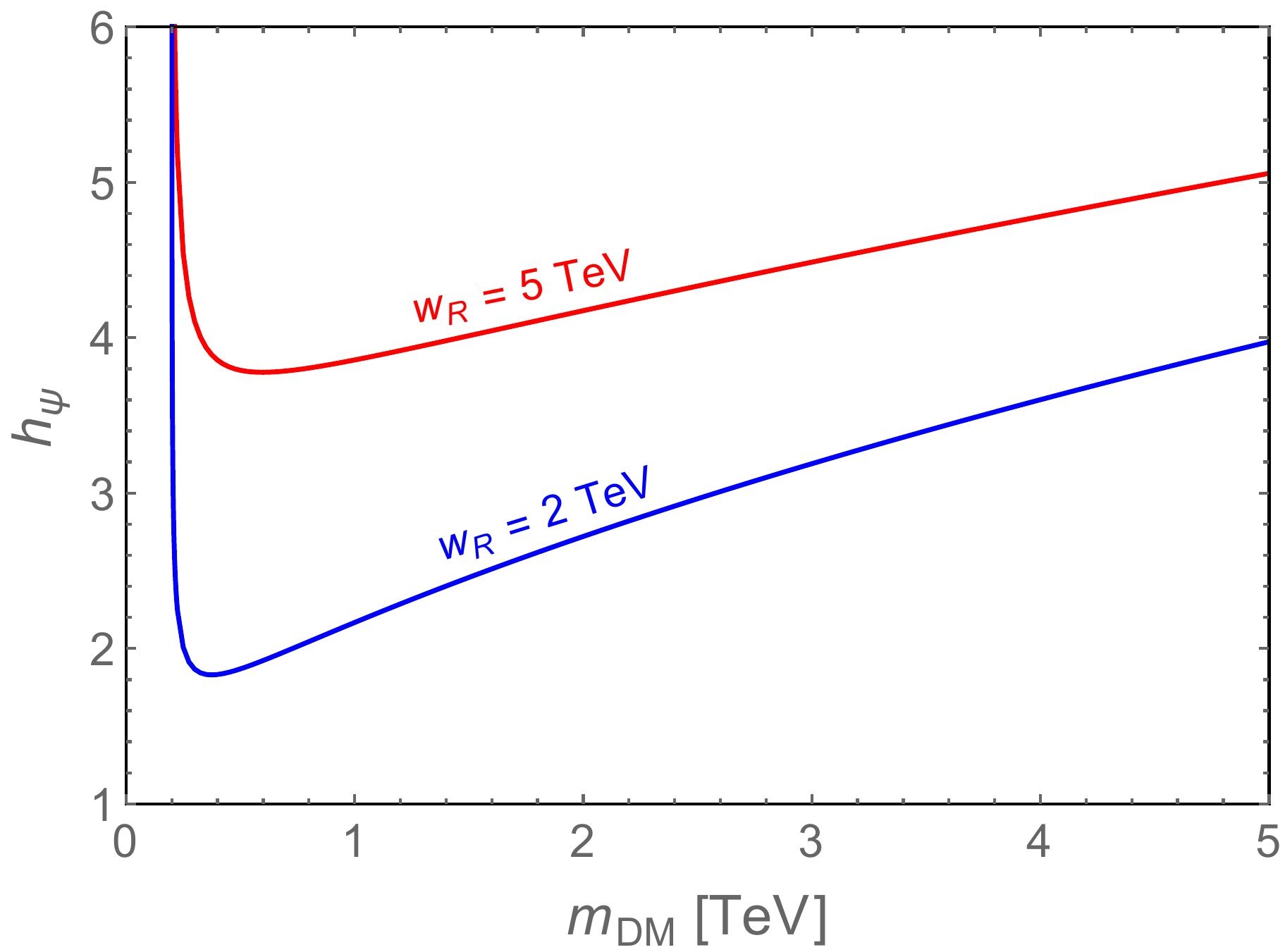}
  \caption{Contours of DM mass $m_{\rm DM}$ and Yukawa coupling $h_\psi$ in Eq.~(\ref{eqn:Yukawa:DM}), which produces the observed DM relic density, for $w_R = 2$ TeV (blue) and 5 TeV (red). We have set $m_\delta = 200$ GeV.}
  \label{fig:DM}
\end{figure}

The resultant contours $m_{\rm DM}$ and $h_\psi$ are shown in Fig~\ref{fig:DM}, which produces the observed DM relic density $\Omega_{\rm DM} h^2 \simeq 0.12$, for two benchmark values of the VEV $w_R = 2$ TeV (blue) and 5 TeV (red). For the sake of concreteness, we have set explicitly the mass $m_\delta = 200$ GeV. Obviously, to have a viable DM from the triplet $\Psi$ in the extended LRSM, the Yukawa coupling in Eq.~(\ref{eqn:Yukawa:DM}) has to be large, of order one, depending on the VEV $w_R$. When the $w_R$ is large, the charged component $\psi^\pm$ is heavier, thus we need a larger Yukawa coupling $h_\psi$ to produce the right DM relic density.

\section{Comments and conclusion}
\label{sec:conclusion}

The LRSM is one of the most economical extensions of the SM to accommodate the tiny neutrino masses via seesaw mechanisms at the TeV scale. However, in the minimal version of LRSM with only a bidoublet $\Phi$ and right-handed triplet $\Delta_R$ in the scalar sector, the neutral scalars $H_1$ and $A_1$ from $\Phi$ mediate tree-level flavor changing processes, which pushes their masses superhigh, approximately greater than $(15 - 25)$ TeV. Consequently, the quartic coupling $\alpha_3 \simeq M_{H_1}^2/v_R^2 \gtrsim {\cal O} (1)$ making it go non-perturbative for a few-TeV  $v_R$ so that the $W_R$ boson is observable at the LHC. This is the FCNH tension for minimal LRSM.

In this paper we have proposed a simple way to ameliorate the FCNH tension in the minimal LRSM, by adding a $B-L=0$ right-handed real triplet $\delta_R$ to the scalar sector. Then the heavy scalars $H_1$, $A_1$ and $H_1^\pm$ from the bidoublet $\Phi$ acquire masses from the trilinear scalar couplings $M_2 {\rm Tr}(\Phi \delta_R \Phi^\dagger)$ which is absent in the minimal LRSM. In the extended LRSM, all the quartic couplings, including $\alpha_3$, remain perturbative at the TeV scale.

Thanks to the new scalar triplet $\delta_R$, the phenomenology in the extended LRSM is very rich:
\begin{itemize}
  \item The addition of the triplet $\delta_R$, while alleviating the FCNH constraints, still leads to its own constraints arising from one loop vacuum stability conditions. These limits imply that for $g_L=g_R$, there is a lower limit on $W_R$ mass of either $4.85$ TeV or $8.1$ TeV depending on whether the LRSM is invariant under parity or under generalized charge conjugation, and the lower limits are respectively 3.16 TeV and 5.26 TeV for the case of $g_R/g_L = 0.65$. These provide a range of masses that are all accessible at the HL-LHC. Clearly, discovery of $W_R$ with mass below these lower limits will rule out this model.

  \item The singly-charged scalar $H_2^\pm$ which is linear combination of $\Delta_R^\pm$ and $\delta_R^\pm$ could mimic the ``smoking-gun'' signal from $W_R$ decay, i.e. $H_2^\pm \to \ell_\alpha^\pm \ell_\beta^\pm jj$, therefore, even if the same-sign dilepton signal is observed at LHC or future colliders, it is {\it not} necessarily from the $W_R$ boson. Our model in this paper provides a good scalar candidate $H_2^\pm$ for the same-sign dilepton signatures. One should, however, note that the production channels of $H_2^\pm$ is very different from the $W_R$ boson, as seen in Section~\ref{sec:singlychargedscalar}.

  \item An interesting aspect of the extended LRSM is a new decay mode of $W_R$ to two scalars, i.e. $W_R \to H_3 H_2^\pm$, that is absent in the minimal LRSM. The branching ratio for this mode is, however, comparatively small, being about $\sim 4\%$, although it may be observable with large statistics.

  \item This model has also interesting implications for several DM models that use $B-L=0$ fermion triplet within the left-right framework. The couplings of $\delta_R$ to DM multiplets could split the masses in the DM sector, whereas in the minimal LRSM all the components in the DM sector are mass degenerate at the tree-level. Even though some simple DM multiplets have been excluded or high constrained in the minimal LRSM, the viable parameter space in this extended LRSM is much larger, as exemplified in Section~\ref{sec:DM}. The same situation happens also if the DM is part of bidoublet fermion. We do not pursue the bidoublet DM model here.
\end{itemize}

We close with the following additional comments:
\begin{itemize}
  \item A question to ask about the model is whether there are any new sources of flavor changing effects due to mixing of the $\delta^0_R$ with the components of the bidoublet scalar. We can illustrate this by looking at the benchmark model, where we note that such mixings in general do exist but as is clear also from the elements in Eqs.~(\ref{eqn:scalarmatrix}) and (\ref{eqn:scalarmixing}), the new FCNH effect induced by the new scalar $H_4$ is down by several orders of magnitude compared to observations.

\item Also our model predicts that the mixing between the SM Higgs boson and the neutral scalar field $\delta_R^0$ to be less than $\sim 10^{-2}$ and therefore any evidence for SM Higgs mixing to a beyond SM scalar larger than this value would rule the model out or at least exclude a large part of the interesting parameter space of the model.

\item While we have discussed how to ameliorate the FCNH tension of type-I seesaw models in this paper, our method can be easily extended to inverse seesaw LRSMs as well~\cite{Dev:2009aw, LalAwasthi:2011aa, Brdar:2018sbk}. We add the same $B-L=0$ scalar field $\delta_R$ to the model and add to the scalar potential the term $M_3\chi^\dagger_R\delta_R\chi_R$ in place of the $M_3 {\rm Tr} (\Delta_R\delta_R\Delta_R^\dagger)$ term in Eq.~(\ref{eqn:potential}). Our discussion in this paper then goes through. Since this is straightforward we do not elaborate on it in this paper. Collider implications on the inverse seesaw LRSMs will of course be different and we do not discuss it here.

\end{itemize}

\section*{Acknowledgements}
The work of R.N.M. and G.Y. is supported by the NSF grant No. PHY1620074, and the work of Y.Z. is supported
by the US Department of Energy under Grant No. DE-SC0017987. R.N.M. would like to thank Jack Collins for some discussions about the model. Y.Z. is grateful to the HKUST Jockey Club Institute for Advanced Study, Hong Kong University of Science and Technology and the Center for High Energy Physics, Peking University for generous hospitality where part of the work was done.

\appendix

%

\section{Partial decay widths of $H_2^\pm$ and $W_R$ }
\label{sec:decay}

The partial decay widths of $H_2^\pm$ are
\begin{eqnarray}
\Gamma (H_2^\pm \to q\bar{q}) \ & = & \
\frac{3\sqrt2 G_F \sin^2 \theta_{\pm} M_{\pm}}{8\pi} \sum_{ij}
\left( |(c_L)_{ij}|^2 + |(c_R)_{ij}|^2 \right) \nonumber \\
&& \times \lambda^{3/2} (m_{u,i}^2,\, m_{d,j}^2,\, M_{\pm}^2) \,, \\
\Gamma (H_2^\pm \to \ell_\alpha^\pm N_\beta) \ & = & \
\frac{\sin^2\varphi_\pm |(f_R)_{\alpha\beta}|^2 M_\pm}{8\pi}
\lambda^{3/2} (0,\, m_{N,\beta}^2,\, M_{\pm}^2) \,,
\end{eqnarray}
where $i,\,j$ are the generation indices for the SM quarks, $\alpha,\,\beta$ are the flavor indices for the leptons, $G_F$ is the Fermi constant, $\sin\theta_\pm$ is the mixing between $H_1^\pm$ and $H_2^\pm$, $f_R$ is the coupling matrix in Eq.~(\ref{eqn:Yukawa}), $\lambda (x,y,z) \equiv (1-x/z-y/z)^2 - 4xy/z^2$, and the coupling coefficients~\cite{Zhang:2007da}
\begin{eqnarray}
c_L \ & = & \ V_{\rm CKM}^{(R)} \widehat{M}_d - 2 \xi \widehat{M}_u V_{\rm CKM}^{(L)} \,, \nonumber \\
c_R \ & = & \ \widehat{M}_u V_{\rm CKM}^{(R)} - 2 \xi V_{\rm CKM}^{(L)} \widehat{M}_d \,,
\end{eqnarray}
with $\widehat{M}_{u,d}$ the diagonal mass matrices for the up and down-type quarks.

In the extended LRSM with the extra triplet $\delta_R$, the partial widths of $W_R$ decays read
\begin{eqnarray}
\Gamma (W_R \to q\bar{q}) & \ = \ &
9\times \frac{g_R^2 M_{W_R}}{48\pi} \,, \\
\Gamma (W_R \to \ell N) & \ = \ &
3\times \frac{g_R^2 M_{W_R}}{48\pi} \,, \\
\Gamma (W_R \to WZ) & \ = \ &
\frac{g_R^2 M_{W_R}}{48\pi} \times \xi^2 \,, \\
\Gamma (W_R \to Wh) & \ = \ &
\frac{g_R^2 M_{W_R}}{48\pi} \times \xi^2 \,, \\
\Gamma (W_R \to WH_3) & \ = \ &
\frac{g_R^2 M_{W}^2}{24\pi M_{W_R}}
\left(\xi^2 \frac{g_R^2}{g_L^2}\right)
\left( 1-x_1 \right)^3
\times \frac{(\cos\theta_5-x\sin\theta_5)^2}{1+2x^2} \,, \\
\Gamma (W_R \to H_2^\pm h) & \ = \ &
\frac{g_R^2 M_{W_R} \varepsilon^2}{48\pi}
\left( 1-x_2 \right)^3
\times (\sin\theta_3-\sin\varphi_2)^2 \,, \\
\label{eqn:WRdecay7}
\Gamma (W_R \to H_2^\pm H_3) & \ = \ &
\frac{g_R^2 M_{W_R}}{48\pi}
\sin^2(\theta_5 + \varphi_\pm) \nonumber \\
&& \times \beta_2 \left[
\left( \beta_2^2 \beta_3 - x_1 - x_2 \right) + \frac14 \beta_2^2 (a_1 - a_2)^2 \right] \,,
\end{eqnarray}
where we have defined $x_1 \equiv M_{H_3}^2 / M_{W_R}^2$, $x_2 \equiv M_{\pm}^2 / M_{W_R}^2$, $\beta_3 = \frac12 (1+a_1a_2)$ and
\begin{eqnarray}
\beta_2 & \ = \ & \sqrt{ 1 - 2 (x_1 + x_2)
+ (x_1 - x_2)^2 } \,, \\
a_1 & \ = \ &
\sqrt{ 1+ 4x_1\beta_2^{-2} } \,, \quad
a_2 \ = \
\sqrt{ 1+ 4x_2\beta_2^{-2} } \,.
\end{eqnarray}

\section{DM annihilation cross section}
\label{sec:DM:cs}

The DM $\psi^0$ annihilates mainly through the process $\psi^0 \psi^0 \to \delta_R^+ \delta_R^-$, with the cross section
\begin{eqnarray}
\sigma (\psi^0 \psi^0 \to \delta_R^+ \delta_R^-) \ = \
\frac{h_\psi^4}{64\pi^2 s} \left( \frac{s-4m_\delta^2}{s-4m_{\rm DM}^2} \right)^{1/2}
\int {\rm d}\Omega \,
({\cal A}_{tt} + {\cal A}_{tu} + {\cal A}_{uu}) \,,
\end{eqnarray}
where
\begin{eqnarray}
{\cal A}_{tt} & \ = \ & \frac{1}{2(t-m_\pm^2)^2} \left[ -3 m_{\rm DM}^4 + 8 m_{\rm DM}^3 m_{\pm} + m_\delta^4+ m_\pm^2 s + tu  -2 m_{\rm DM} m_\pm (s - t + u) \right. \nonumber \\
&& \left. + m_{\rm DM}^2 \left( -4 m_\pm^2 + 2 m_\delta^2 + s - 3t + u  \right)
- m_\delta^2 (s+t+u) \right] \,, \\
{\cal A}_{tu} & \ = \ & \frac{1}{2(t-m^2_\pm)(u-m^2_\pm)} \left[ -6 m_{\rm DM}^4 + 16 m_{\rm DM}^3 m_{\pm} + 2 m_\delta^4 - 4 m_{\rm DM} m_\pm s  + 2 m_\pm^2 s \right. \nonumber \\
&& \left. - (s^2 -t^2 -u^2)  + 2 m_\delta^2 (s - t - u)
- 2 m_{\rm DM}^2 (4 m_\pm^2 - 2 m_\delta^2  - s + t + u) \right] \,, \\
{\cal A}_{uu} & \ = \ & \frac{1}{2(u-m^2_\pm)^2} \left[ -3 m_{\rm DM}^4 + 8 m_{\rm DM}^3 m_{\pm} + m_\delta^4+ m_\pm^2 s + tu  -2 m_{\rm DM} m_\pm (s + t - u) \right. \nonumber \\
&& \left. + m_{\rm DM}^2 \left( -4 m_\pm^2 + 2 m_\delta^2 + s + t -3 u  \right)
- m_\delta^2 (s+t+u) \right] \,.
\end{eqnarray}


\end{document}